\documentclass[aps,prx,twocolumn,superscriptaddress,longbibliography]{revtex4-2}
\usepackage{color}
\usepackage{dcolumn}
\usepackage{graphicx}
\usepackage{amsmath}
\usepackage{amssymb}
\usepackage{bm}
\usepackage{braket}
\usepackage[hidelinks]{hyperref}
\usepackage{multirow}
\usepackage{ulem}

\usepackage{xcolor} 
\usepackage{todonotes} 
\usepackage{soul} 
\definecolor{red}{RGB}{220, 20, 60}
\definecolor{darkgreen}{RGB}{0, 100, 0}

\usepackage{soul}
\usepackage{xcolor}

\begin{document}

\title{Mitigating state transition errors during readout with a synchronized flux pulse}
\author{Yulong Li}
\thanks{Equal contribution author}
\affiliation{Beijing Key Laboratory of Fault-Tolerant Quantum Computing, Beijing Academy of Quantum Information Sciences, Beijing, China, 100193}
\author{Wuerkaixi Nuerbolati}
\thanks{Equal contribution author}
\affiliation{Beijing Key Laboratory of Fault-Tolerant Quantum Computing, Beijing Academy of Quantum Information Sciences, Beijing, China, 100193}

\author{Chunqing Deng}
\affiliation{Quantum Science Center of Guangdong-Hong Kong-Macao Greater Bay Area, Shenzhen, China}

\author{Xizheng Ma}\email{maxizheng@quantumsc.cn}
\affiliation{Quantum Science Center of Guangdong-Hong Kong-Macao Greater Bay Area, Shenzhen, China}

\author{Haonan Xiong}\email{xionghn13@gmail.com}
\affiliation{Beijing Key Laboratory of Fault-Tolerant Quantum Computing, Beijing Academy of Quantum Information Sciences, Beijing, China, 100193}

\author{Haifeng Yu}
\affiliation{Beijing Key Laboratory of Fault-Tolerant Quantum Computing, Beijing Academy of Quantum Information Sciences, Beijing, China, 100193}
\affiliation{Hefei National Laboratory, Hefei, China, 230088}

\date{\today}

\begin{abstract}


State transitions during qubit measurements are extremely detrimental to quantum tasks that rely on repeated measurements, such as quantum error correction. These state transitions can occur when excessive measurement power leads to qubit excitations outside its computational space. Alternatively, the qubit state can decay rapidly when the measurement protocol inadvertently couples the qubit to lossy modes such as two-level systems (TLSs). We experimentally verify the impact of these TLSs in qubit readout by measuring the transition errors at different qubit flux bias. Because such state transitions during measurements are often localized in frequency space, we demonstrate the ability to avoid them during a fluxonium readout by exploiting the qubit's flux-tunability. By synchronizing the flux bias with the readout photon dynamics, we obtain an optimal readout fidelity of 99 \% (98.4 \%) in 1 $\mathrm{\mu s}$ (0.5 $\mathrm{\mu s}$) integration time. Our work advances the understanding of state transitions in superconducting circuit measurements and demonstrates the potential of fluxonium qubits to achieve fast high-fidelity readout.
\end{abstract}

\maketitle

\section{Introduction}
High-fidelity quantum non-demolition (QND) readout is an essential ingredient for quantum information processing. In superconducting circuits, dispersive measurement~\cite{Blais_review} is the method of choice for state discrimination, enabling remarkably fast and accurate assignment of qubit states. Indeed, $>99\%$ readout fidelities have been reported for measurements performed under $100~\text{ns}$ in transmon systems~\cite{Sunada2022,Sunada2024,Walter2017,Swiadek2024}, while advances have also been recently reported for the less explored fluxonium qubits~\cite{stefanski_improved_2024, Bothara2025}. Meanwhile, for many quantum protocols relying on repeated qubit measurements -- such as erasure conversion~\cite{Levine2024,Koottandavida2024} and quantum error correction~\cite{Google2023,Ni2023} -- a high degree of QNDness is required, where the post-measurement qubit state should remain the same as the previous measurement result. Therefore, suppressing non-QND state transitions while maintaining a highly accurate state assignment is crucial, but remains challenging.

A trade-off often exists in measurement between the signal-to-noise ratio (SNR) and QNDness. While increasing the readout power offers a straightforward approach to enhancing SNR, excessive power risks exciting the qubit beyond its computational space~\cite{Sank2016,Khezri2023,Bothara2025,Bista2025,Hazra2025, Wang2025,dai_spectroscopy_2025}. Recent theoretical work shows that these so-called measurement-induced-state-transitions (MIST) likely arise from energy exchange between non-computational qubit states and excitations in either the readout resonator~\cite{Shillito2022,Nesterov2024} or spurious circuit modes~\cite{Singh2024}. This effect complicates the readout optimization especially for highly anharmonic systems such as fluxoniums. Alternatively, a careful balance of the resonator dispersive shift $\chi$, the photon loss rate $\kappa$, and the measurement speed can be performed to optimize the readout SNR~\cite{Walter2017}. Recent attempts~\cite{Swiadek2024,Stefanski2024,stefanski_improved_2024} have significantly reduced design constraints by enabling in-situ adjustment of $\chi$ through an adiabatic variation of the qubit frequency, and consequently the qubit-resonator detuning, via a dynamical flux pulse. Effectively separating the qubit's gate operation position from its readout position, this technique potentially has the combined benefit of optimizing the measurement SNR with optimal $\chi$ while protecting the qubit from residual resonator photons~\cite{clerk_using_2007,rigetti_superconducting_2012,zhang_suppression_2017,wang_cavity_2019} with a minimal $\chi$ during gate operations. However, by changing the qubit frequency, the flux pulse risks bringing the qubit into resonance with two-level-systems (TLSs)~\cite{barends_coherent_2013} residing along the modulation path, thereby inducing state transitions during measurements~\cite{Sank2016,thorbeck_readout-induced_2024,Bista2025, Hazra2025}.
There is still a lack of research and mitigation strategies targeting these issues.


In this work, we conduct a comprehensive exploration of the readout parameter space for a typical fluxonium qubit~\cite{manucharyan_fluxonium_2009}, and optimize its readout fidelity. Leveraging the flux-tunability of the fluxonium qubit, we adjust the qubit flux bias during readout and characterize the dependence of the state transition probability on the qubit energy landscape, average photon number, and qubit state. These measurements confirm two origins for state transitions during measurement: MIST effects and spurious TLSs. To mitigate TLS-induced state transitions, we synchronize the qubit flux bias with the photon dynamics in the readout resonator to ensure an optimal path in the readout flux-power parameter space. Finally, we balance readout assignment error with state transition errors to achieve a high readout fidelity of 99.0~\% within 1~$\mathrm{\mu s}$ or 98.4~\% within 0.5~$\mathrm{\mu s}$. Our readout fidelity and gate speed are at the forefront in the field compared with recent fluxonium experiments~\cite{ding_high-fidelity_2023,Bothara2025,Bista2025,xiong_scalable_2025} despite the strongly-coupled TLSs near the half-integer flux.

\section{System parameters}
\begin{figure}[h]
    \centering
    \includegraphics[width=1.0\linewidth]{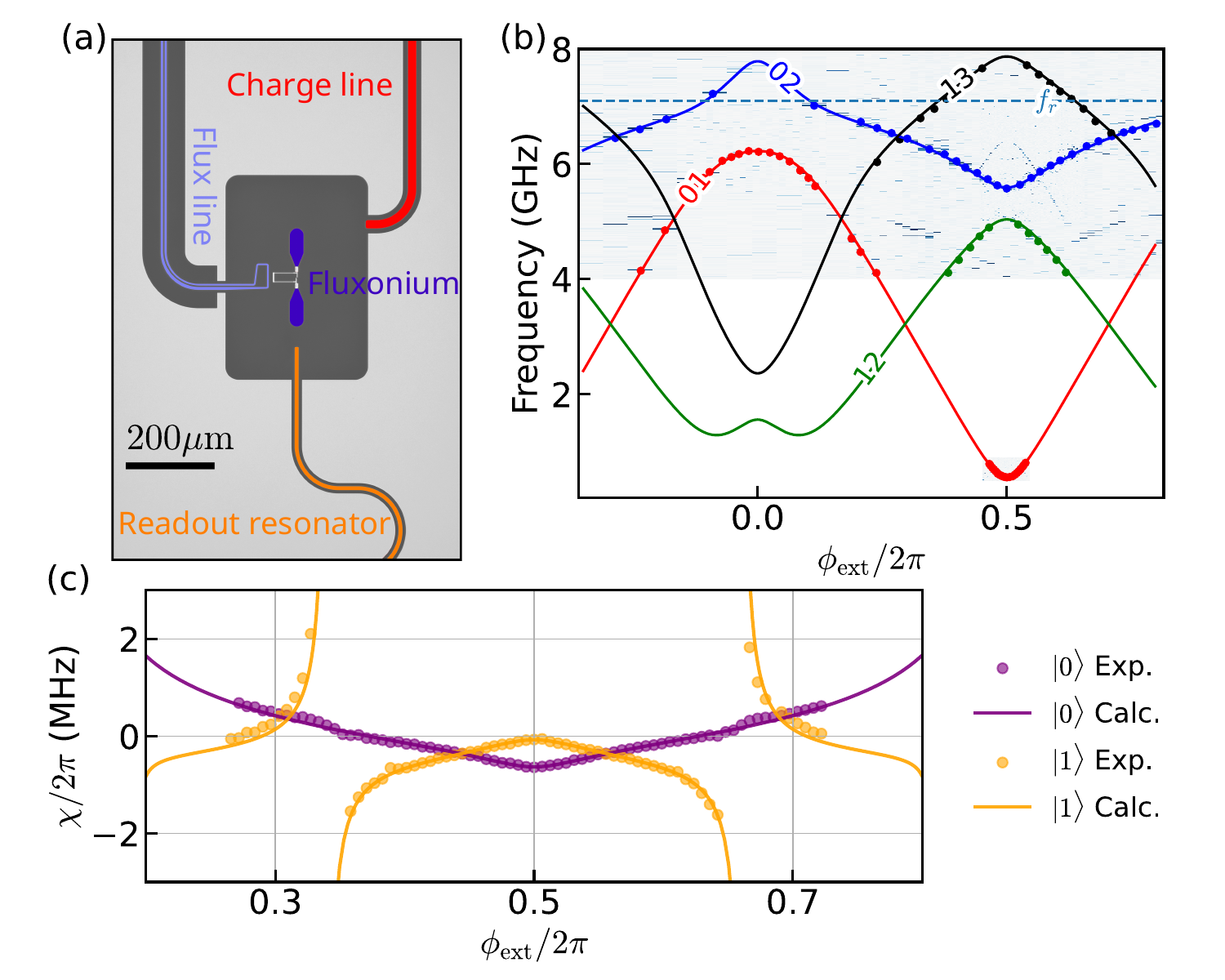}
    \caption{The measured device. (a) False-color optical microscope image of a fluxonium qubit (dark blue) coupled to a charge line (red), a flux line (light blue) and a readout resonator (orange). (b) Spectroscopy of the fluxonium qubit as a function of external flux. (c) Dispersive shift $\chi$/2$\pi$ as a function of flux with qubit prepared either at $|0\rangle$ or $|1\rangle$ states. The dots and lines respectively correspond to the measurement and theory.}
    \label{fig:fig1}
\end{figure}

Our circuit consists of a single fluxonium qubit capacitively coupled to a readout CPW resonator, as shown in Fig.~\ref{fig:fig1}(a).
The resonator is coupled to a Purcell filter of 150-MHz bandwidth (not shown). The system is described by the Hamiltonian 
\begin{equation}\label{eq:H}
    \begin{aligned}
        H/h = &4E_{C}\hat{n}^2+\frac{1}{2}E_{L}\hat{\varphi}^2
        -E_{J} \cos(\hat{\varphi}-\phi_{\rm ext})+f_r\hat{a}^\dag \hat{a}\\
        +&g_{na}\hat{n}(\hat{a}+\hat{a}^\dag),
    \end{aligned}
\end{equation}
where $E_C = 1.848$ GHz, $E_J = 4.684$ GHz and $E_L = 0.491$ GHz are extracted from the measured qubit spectrum in Fig.~\ref{fig:fig1}(b). In Fig.~\ref{fig:fig1}(c), we extract the dispersive shift as a function of flux by measuring the resonator spectrum for different qubit state. From this measurement, we obtain the bare resonator frequency $f_r = 7.105$ GHz and the qubit-resonator coupling strength $g_{na} = 53.7$ MHz. At the half-integer flux, henceforth referred to as the sweet spot, the qubit frequency is 564 MHz and the dispersive shift $\chi/2\pi = 0.57$ MHz. The qubit initialization is performed by appending a $\pi$- or identity-pulse to a sideband reset~\cite{wang_efficient_2024}. 
We also measure the resonator linewidth $\kappa/2\pi = 3.50$ MHz from resonator photon dynamics (Appendix~\ref{sec:flux_cal}). 

\section{state transitions during measurement}
\begin{figure}[h]
    \centering
    \includegraphics[width=1.0\linewidth]{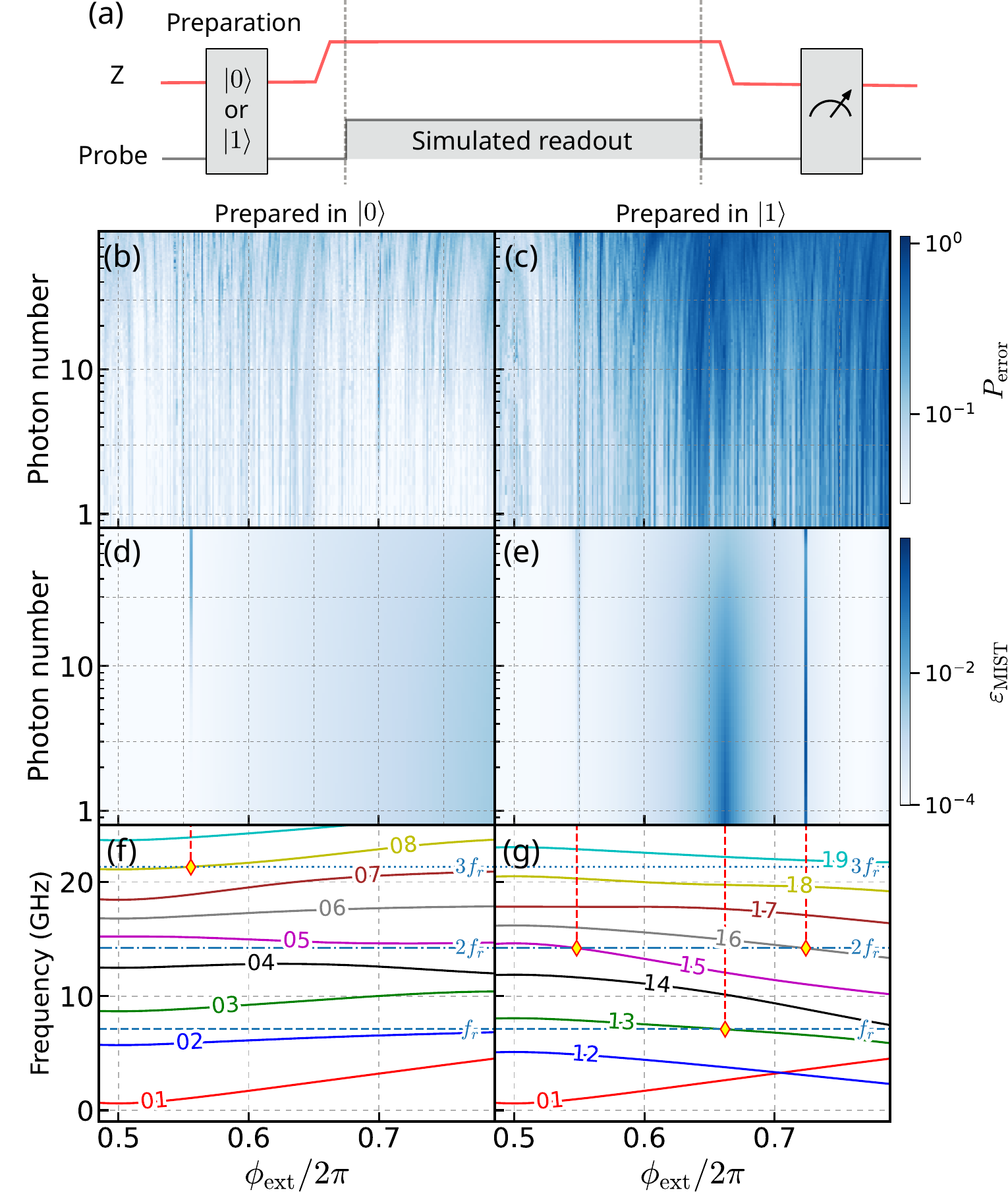}
    \caption{(a) Pulse sequence for readout back-action measurement. To simulate the effect of a readout pulse, we apply a $2~\mathrm{\mu s}$ microwave pulse on the resonator. An accompanying flux pulse is also applied to simulate readouts at different bias locations. The state transition error $P_\mathrm{error}$ is defined as the probability $P(1|0)$ of measuring $\ket{1}$ when the qubit is initialized in $\ket{0}$ (b), or the probability $P(0|1)$ of measuring $\ket{0}$ when the qubit is initialized in $\ket{1}$ (c). (d,e) One major source of state transitions error arises from MIST events, which we predict using the error metric $\epsilon_\text{MIST}$ that quantifies how much the dressed-coherent-state picture of the qubit-resonator system would break when one additional readout photon is added or removed from the resonator. (f,g) These MIST errors arise when the multiple readout photons have the correct amount of energy to excite the fluxonium to a high-energy state. The red dashed lines and yellow diamond markers indicate locations where such frequency collisions occur. 
    }
    \label{fig:fig2}
\end{figure}

To investigate the readout back-action on the fluxonium qubit, we measure the state transition error $P_\mathrm{error} = P(0|1)$ for qubit initialized in $\ket{1}$ or $P_\mathrm{error} = P(1|0)$ for states initialized in $\ket{0}$, as a function of both qubit flux bias and resonator photon number during readout. As shown in Fig.~\ref{fig:fig2}(a), this is done by applying a $2~\mathrm{\mu s}$ microwave pulse of varying power to the resonator to simulate the effect of different readout powers, whereas an accompanying flux pulse of varying amplitude is used to simulate readouts at different bias locations. A final measurement of fixed power and flux then quantifies the impact of the simulated readout on the prepared qubit state. The observed MIST errors for the qubit initialized at $|0\rangle$ and $|1\rangle$ states are shown in Fig.~\ref{fig:fig2}(b) and (c), respectively. For different flux biases, we ensure the readout simulation pulse is always on resonance with the qubit-state-dressed resonator, and the resonator photon number is calibrated through independent ac Stark shift measurements.

One source for these state transitions is MIST events, which could happen when multiple readout photons have the correct amount of energy to excite the fluxonium to a high-energy state. Following Ref.~\cite{Nesterov2024}, we predict the onset of these transitions by calculating the error metric $\epsilon_\text{MIST}$ that quantifies how much the dressed-coherent-state picture of the qubit-resonator system would break when one additional readout photon is added or removed (see Appendix~\ref{sec:mist_err}). Indeed, Fig.~\ref{fig:fig2}(d,e) reveals four clusters of MIST errors, corresponding to locations where the transition energies of fluxonium 0-8, 1-5, 1-3, and 1-6 are integer multiples of the readout photon energy (Fig.~\ref{fig:fig2}(f,g)). Of these four MIST processes, we observe clear agreement between measurement and theory for the 1-5 and 1-3 transitions, but the MIST transitions for 0-8 and 1-6 are much harder to confirm. Multiple factors could contribute to this. First, we predict high-energy fluxonium transitions, and hence the location of MIST events, based on qubit spectrum measured below 8~GHz, which can be markedly inaccurate at higher frequencies due to the existence of junction array modes \cite{kuzmin_superstrong_2019,mehta_quantum_2022, mencia_ultra-high_2023}. Second, because our final measurement performs a binary assignment of the measurement IQ signals to $\ket{0}$ and $\ket{1}$, we may miss state transitions to higher-energy states when their signals overlap significantly with that of the ``correct" state~\cite{Hazra2025}. Finally, the abundance of state transition events obscures genuine MIST events, making them difficult to distinguish from other error sources --- particularly frequency collisions between the qubit and parasitic TLSs that accelerate qubit decay and excitation during measurement.

\section{Mitigating transition errors induced by two-level systems}
\begin{figure}[h]
    \centering
    \includegraphics[width=0.9\linewidth]{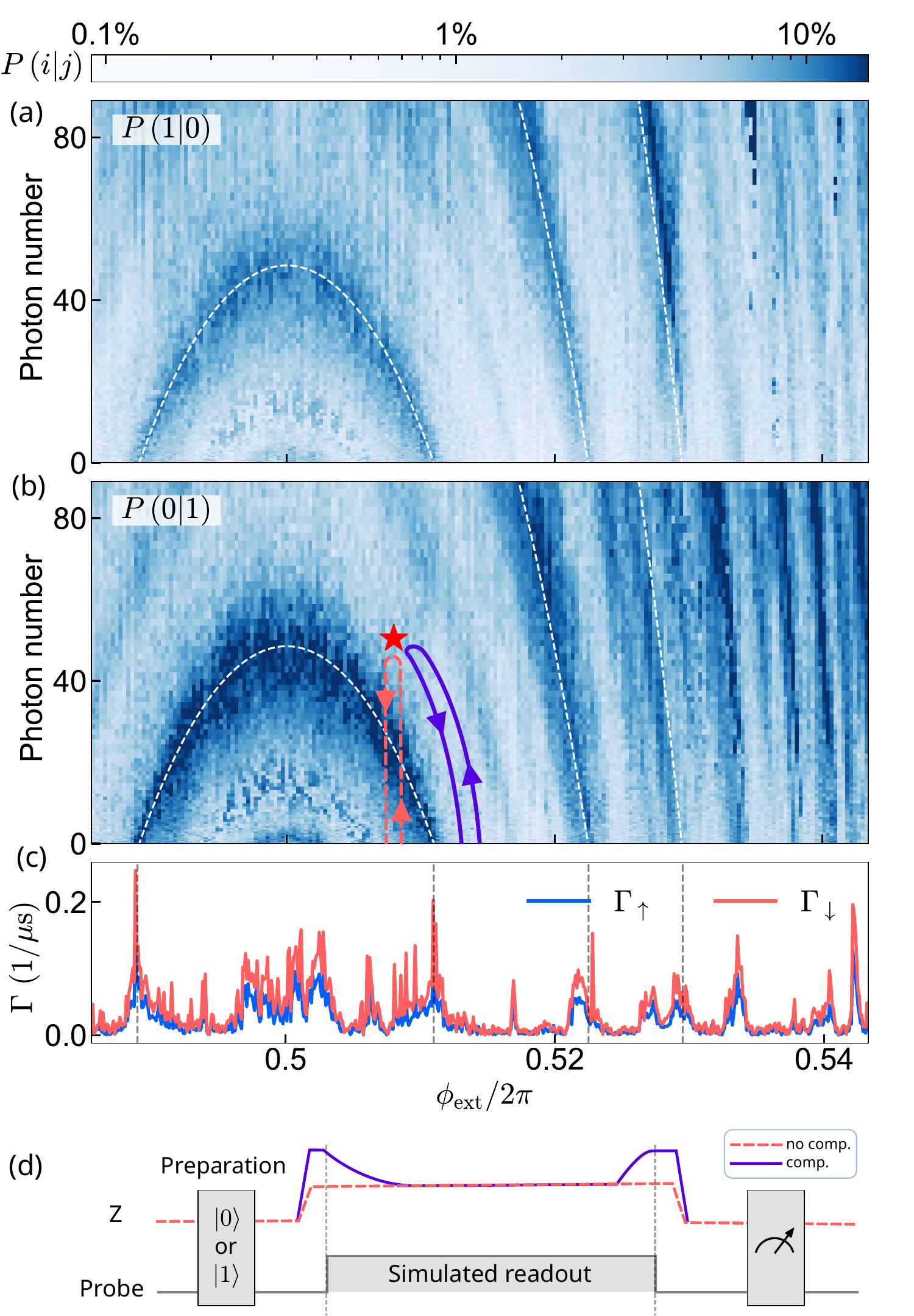}
    \caption{
    (a,b) TLS-induced state transition errors are particularly prominent around the half-flux quantum. The fringe patterns are caused by resonance conditions between fixed-frequency TLSs and the ac Stark shifted qubit frequency. (c) The separately measured qubit depolarization rates as a function of flux bias reveal locations of the TLSs, from which we calculate the white dashed lines in panels (a) and (b) with no free parameters. Crossing these TLS-boundaries during the transient dynamics of the resonator (red trajectory in panel (b)) incurs state transition errors. (d) To combat such errors, we introduce a synchronized flux compensation, which results in the optimized purple trajectory in panel (b).
    }
    \label{fig:fig3}
\end{figure}
To study the transition errors induced by these parasitic TLSs, we perform finer scans near the half-flux quantum in Fig.~\ref{fig:fig3}(a,b). By separate measurements of the qubit lifetime as a function of flux bias, we can obtain the location of TLSs that degrade qubit coherence (Fig.~\ref{fig:fig3}(c)), which we model as fixed in frequency. In our experiment, we often observe double-exponential behavior in qubit relaxation, which can be attributed to a bath of long-lived TLS~\cite{zhuang_non-markovian_2025}. Consequently, we extract the relaxation rate $\Gamma_\downarrow$ and excitation rate $\Gamma_\uparrow$ from the first $2.5~\mathrm{\mu s}$ in a $T_1$ measurement. Because the dispersive shift $\chi(\phi_\mathrm{ext})$ is positive around $\phi_\mathrm{ext}=\pi$, photons in the resonator will Stark shift the qubit a higher frequencies,
\begin{equation}
    f'_{01}(n, \phi_\mathrm{ext}) = f_{01}(\phi_\mathrm{ext}) + n \chi(\phi_\mathrm{ext}),
    \label{eq:f01}
\end{equation}
where $f_{01}$ is the original qubit frequency.
Consequently, the hyperbolic fringes observed in panels (a) and (b) emerge from the flux and photon number dependence of the resonance condition between fixed frequency TLSs and the Stark-shifted qubit frequency $f'_{01}$. Indeed, picking three prominent TLSs in Fig.~\ref{fig:fig3}(c), we plot 
their respective resonance conditions $n(\phi_\mathrm{ext})$ (white dashed lines, Fig.~\ref{fig:fig3}(a,b)). The model reproduces the observed fringe shapes with no fitting parameters, showing excellent agreement. Additionally, the measured relaxation rates $\Gamma_\downarrow$ are typically higher than the excitation rates $\Gamma_\uparrow$, consistent with the higher transition errors when the qubit is prepared in $\ket{1}$ state. 
At larger photon numbers, the increased measurement-induced dephasing causes the broadening of the qubit linewidth~\cite{thorbeck_readout-induced_2024}. This likely results in both the broadening of the fringe features and the increase in background error at larger photon numbers.

The hyperbolic-shaped resonance conditions between TLS and Stark-shifted qubits create enclosed parameter spaces within which qubit readout incurs less transition error. Meanwhile, crossing these resonance boundaries even during the transient dynamics of the resonator would lead to increased non-QND errors. One way to reduce such error is to employ advanced pulse shaping on the resonator drive~\cite{mcclure_rapid_2016,bultink_general_2018,jerger_dispersive_2024,Hazra2025} to reduce the transient time. Alternatively, here we propose to avoid the crossing of boundaries by synchronizing the flux bias with resonator dynamics. Specifically, we shape our flux bias (Fig.~\ref{fig:fig3}(d)) to compensate for the Stark shift on the qubit, such that $f'_{01}$ remains constant (Eq.~\ref{eq:f01}). This results in an optimal trajectory as shown in Fig.~\ref{fig:fig3}(c) with the purple line. The calibration details on the compensation pulse are discussed in Appendix~\ref{sec:flux_ramsey}. Naturally, the qubit still inevitably crosses TLSs when its flux is tuned from the fluxonium sweet spot to the start of the readout trajectory, but this speed does not depend on the resonator linewidth and can be much faster than the photon transient time. 

\begin{figure}[h]
    \centering
    \includegraphics[width=0.9\linewidth]{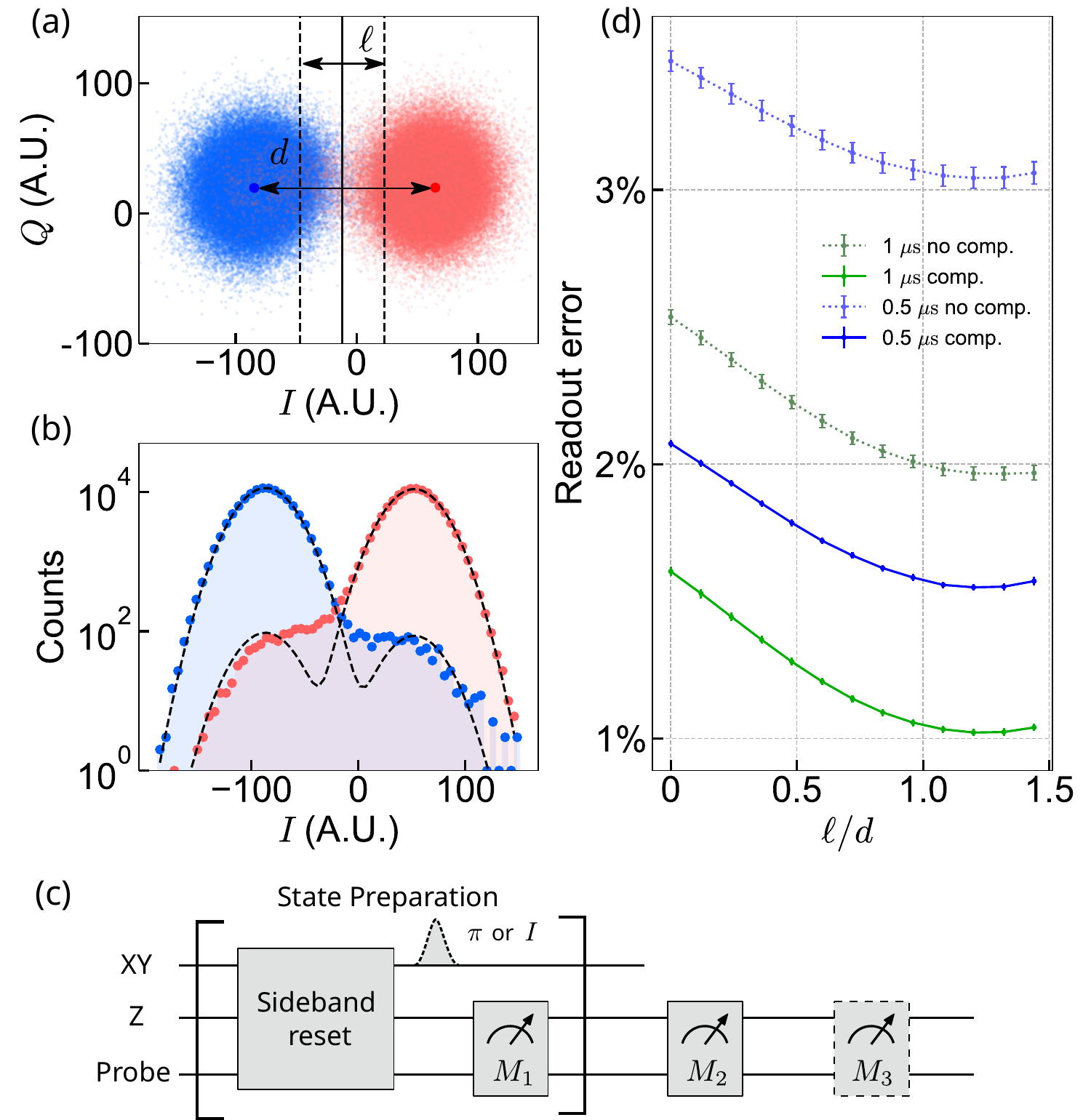}
    \caption{Optimized performance for a $1~\mu$ readout. (a) Post-selected single-shot results with blue (red) points corresponding to $\ket{0}$ ($\ket{1}$) state. The vertical solid line corresponds to the binary assignment threshold. For post-selecting the state preparation, we discard the dataset when the first measurement falls between the dashed lines. (b) Single-shot histograms of the in-phase quadrature. (c) Sequence for readout characterization. The first measurement $M_1$ is for post-selection. The third measurement $M_3$ is for classification of the readout errors in $M_2$. (d) Readout error vs the selection gap $\ell$ for different readout pulse. The error bars are standard error from 50 repeated measurements. 
    }
    \label{fig:fig4}
\end{figure}

\section{Readout optimization}
For a given readout duration, a large SNR requires more photons during the readout, but could cause increased transition error. Consequently, the optimal readout is achieved through a balance between the SNR-limited assignment errors and transition errors. To this end, we perform numerical optimizations (CMA-es) to find the optimal 
 combination of readout power, flux bias and  frequency for 1 $\mathrm{\mu s}$ and 0.5 $\mathrm{\mu s}$ readout pulses respectively with compensation flux control turned on (purple trajectory in Fig.~\ref{fig:fig3}(c)). Fig.~\ref{fig:fig4}(a) shows the measured single-shot result for the optimized 1 $\mathrm{\mu s}$ readout. We then perform a double Gaussian fit on its histogram in Fig.~\ref{fig:fig4}(b). Near the center of the histogram, there is significant deviation between measurement and fit, which we attribute to state transitions during the readout, particularly those within the computational basis.

To distinguish measurement error from state preparation error, we perform an aggressive post-selection on the measurement during the state-preparation stage ($M_1$ in Fig.~\ref{fig:fig4}(c)). Motivated by the non-negligible overlap between the two qubit blobs, we discard all instances where the first measurement falls within a distance of $\pm l/2$ from the binary threshold (Fig.~\ref{fig:fig4}(a)). Indeed, as shown in Fig.~\ref{fig:fig4}(d), the error in measurement $M_2$ is suppressed as we increase $l$ before saturating toward the pure readout error of $1.0\%$ for a 1 $\mathrm{\mu s}$ pulse and $1.6\%$ for a 0.5 $\mathrm{\mu s}$ pulse. Importantly, when we remove the flux compensation (red trajectory in Fig.~\ref{fig:fig3}(c)), the readout errors almost double. This shows that the TLS can induce significant state transition errors during readout. We note that the error bars in Fig.~\ref{fig:fig4}(d) are dominated by long-time fluctuations of the repeated measurements, which is a typical feature of TLSs~\cite{klimov_fluctuations_2018,thorbeck_two-level-system_2023}. 

Finally, to distinguish between transition errors and assignment errors, we add a third readout $M_3$ (Fig.~\ref{fig:fig4}(c)) in our experiment~\cite{kurilovich_high-frequency_2025}. For cases where $M_1 = M_3 \neq M_2$, an assignment error has occurred during $M_2$. Alternatively, when $M_2 = M_3 \neq M_1$, a state-transition error has occurred during $M_2$. For both 1 $\mathrm{\mu s}$ and 0.5 $\mathrm{\mu s}$ readout pulses with flux compensation on, we find 67 \% of the total error comes from state transition and the rest is assignment error.

\section{Summary and Outlook}
In this work, we have investigated state transition errors during fluxonium qubit readout using dynamical flux pulsing and identified MIST and TLS-induced state transitions as the two dominant error sources. To combat TLS-induced errors, we introduce a flux pulse compensation technique to optimize the readout trajectory in the readout flux-power landscape. Finally, by a careful balance between state transition errors and assignment errors, we achieve a pure readout error of 1.0 \% for a 1 $\mathrm{\mu s}$ pulse and 1.6 \% for a 0.5 $\mathrm{\mu s}$ pulse. 

Looking forward, the fluxonium readout still has much room for improvement. 
While the TLS-induced errors can be partially mitigated using the flux-compensation pulse, they eventually impede readout at large photon number. Consequently, improving fabrication to reduce the amount of TLS is immensely important for fluxonium qubits. 
Moreover, in our experiment, we extract~\cite{bultink_general_2018} a measurement efficiency of $\eta=17 \%~(2.4 \%)$ with (without) the phase-preserving Josephson Parametric Amplifier (JPA). Significant improvements can be made by operating the quantum amplifier in the phase-sensitive mode\cite{castellanos-beltran_development_2010,weber_quantum_2014,touzard_gated_2019,Hazra2025} and employing proper weighting functions~\cite{bultink_general_2018}.

\begin{acknowledgements}
We thank Dr. He Wang for the helpful discussions. We acknowledge the support from the National Natural Science Foundation of China (Grant No. 92365206) and the Innovation Program for Quantum Science and Technology (No. 2021ZD0301802). C. Deng and X. Ma also acknowledge the support from Guangdong Provincial Quantum Science Strategic Initiative (Grant No. GDZX2407001).

\end{acknowledgements}

\appendix





\section{Device fabrication \label{sec:fab}}
The fluxonium device is fabricated on a 0.43 mm-thick sapphire substrate where a 200 nm $\alpha$-tantalum film is deposited by DC-sputtering. The circuit structures, including qubit pads, readout resonators, and transmission lines, are patterned via direct writing lithography and are dry etched using a RIE system with CF$_4$ gas. Prior to the fabrication of Josephson junctions, the wafer is soaked in a 2:1 H$_2$SO$_4$:H$_2$O$_2$ piranha solution for 20 mins at room temperature to remove organic residuals on the device surface. A LOR-10B/PMMA A4/AR-PC 5092 trilayer resist is spin-coated on the wafer, followed by e-beam Lithography to pattern the SQUID loop with integrated junction structures. After resist development, the wafer is loaded into a four-chamber JEB system, where 40 nm/85 nm Al-AlO$_x$-Al Josephson junctions and Josephson junction arrays are fabricated by a double-angle evaporation process and an in situ oxidation process to construct the fluxonium qubit.

\section{Experiment Setup \label{sec:setup}}
\begin{figure}[h]
    \centering
    \includegraphics[width=0.9\linewidth]{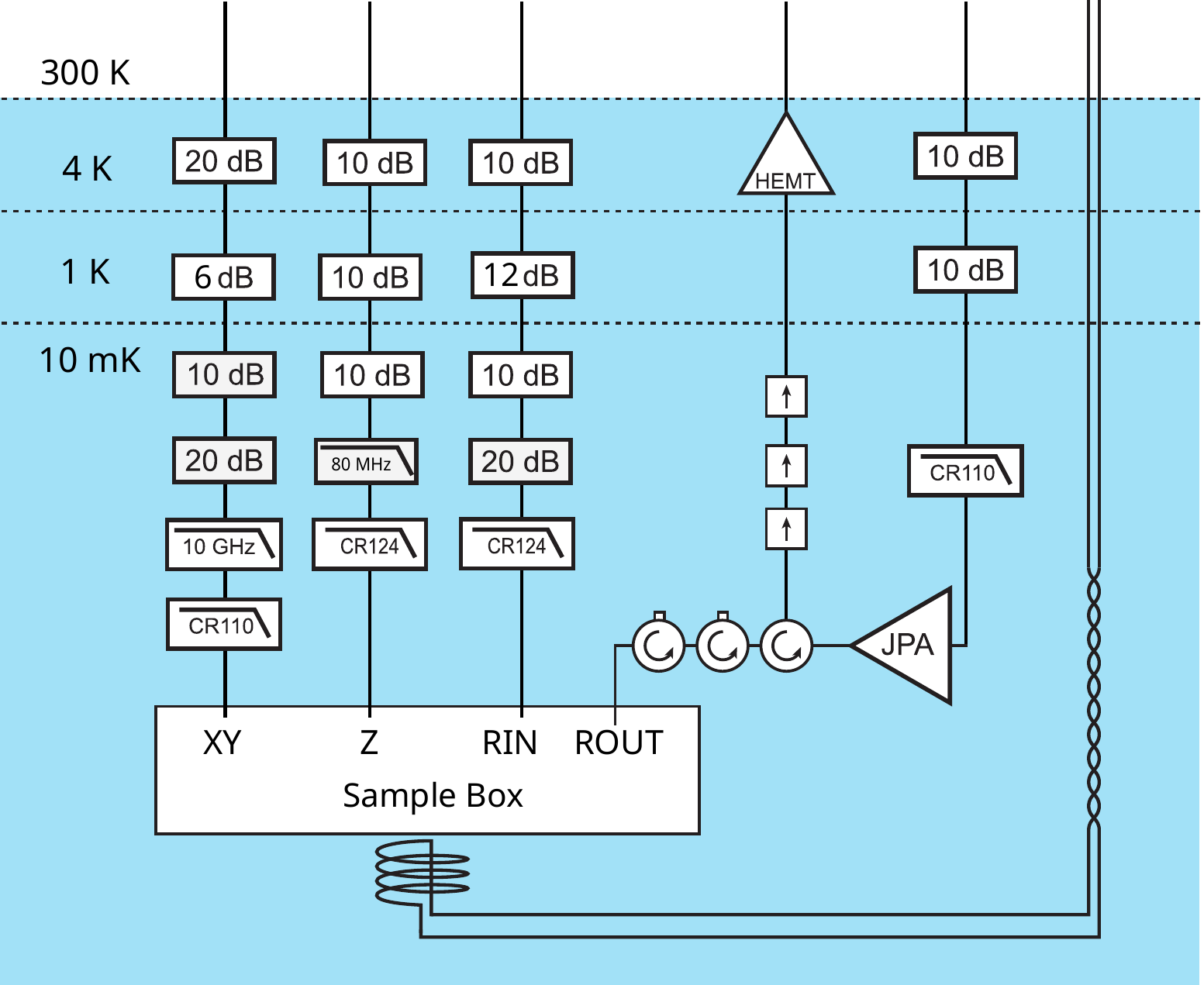}
    \caption{
       Cryogenic wiring of the setup in this experiment.
    }
    \label{fig:fridge}
\end{figure}

The fluxonium sample is wire-bonded to a PCB board and housed in a copper box. A coil made of aluminum wire is mounted to the box to bias the qubit at half-integer flux without heating the fridge. The attenuation and filtering on the control lines are presented in Fig.~\ref{fig:fridge}. The readout signal is directed to a reflective JPA with the bias and pump applied to its flux line. The sample is measured in a Bluefors LD400 dilution fridge. The qubit reset and readout pulses are generated with NAISHU QC110 and QMAC-2120 which also performs demodulation on the output signal from the fridge. The qubit XY and Z pulses are generated with ChipQ-AWG-4. The DC currents are applied with Rigol DG1062. The JPA pump is a continuous wave generated from Keysight N5183B.

\section{Photon dynamics during a readout pulse\label{sec:flux_cal}}
\begin{figure}[h]
    \centering
    \includegraphics[width=\linewidth]{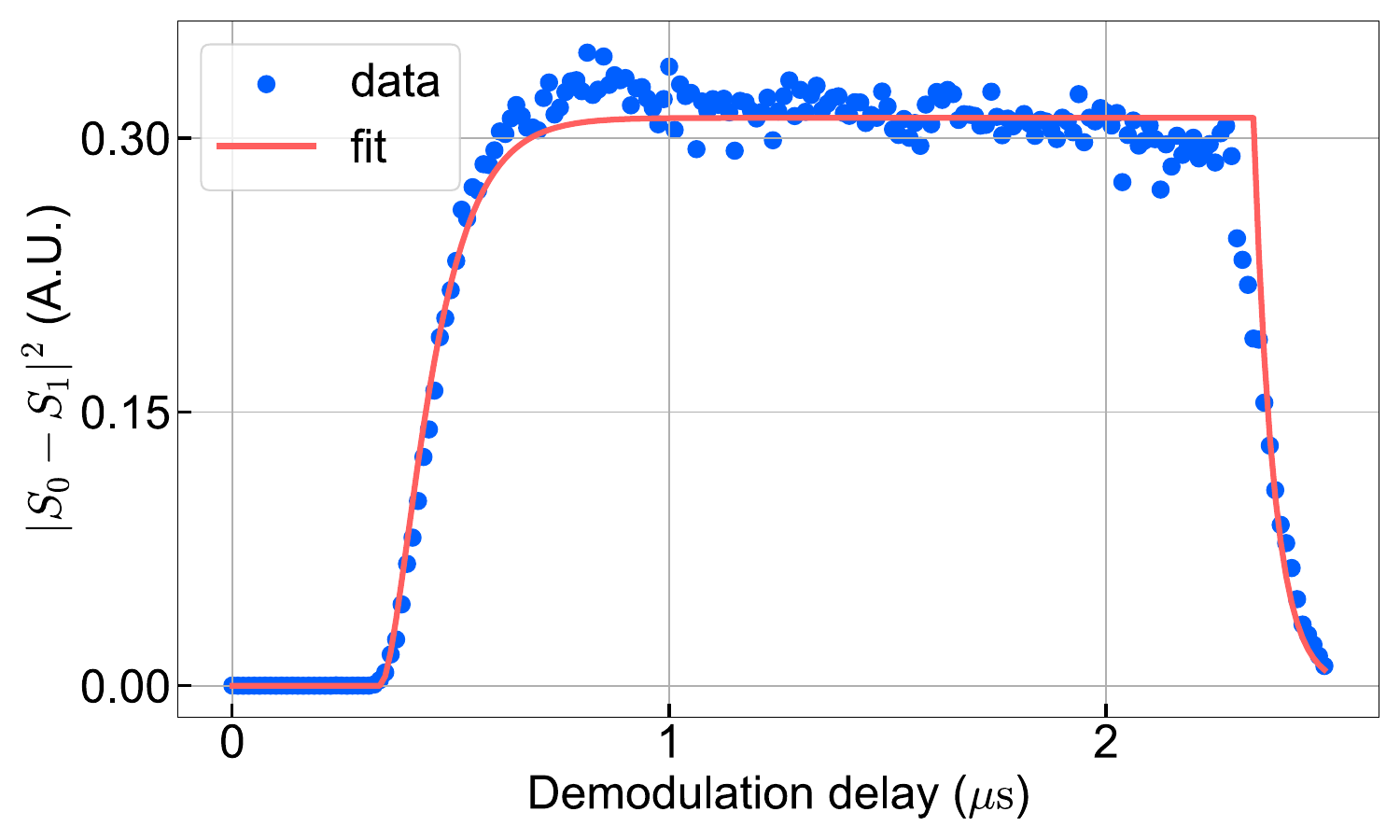}
    \caption{
       Photon dynamics measurement. We use a 10-ns demodulation window and the calculate the signal contrast between the two qubit states. Its squared modulus is proportional to the number of readout photons.
    }
    \label{fig:photontime}
\end{figure}
In our experiment, a square readout pulse with a drive strength $\epsilon$ is applied to the readout resonator. After the beginning of the pulse, the average photon number in the resonator is given by $\langle \hat{n} (t) \rangle  = \bar{n} [ 1 + \exp{(- \kappa t )} - 2\cos(\Delta t) \exp{( - \kappa t  / 2)} ]$, where $\Delta$ is the drive detuning and the steady state photon number $\bar{n}=4|\epsilon|^2/(\kappa^2+4\Delta^2)$. Because $\Delta$ is much smaller than $\kappa$, we use $\langle \hat{n} (t) \rangle  = \bar{n} [ 1 + \exp{(- \kappa t )} - 2 \exp{( - \kappa t  / 2)} ]$ in practice and assume $\bar{n}$ is identical for the two qubit states. When the driving pulse stops, the photon decays exponentially $\langle \hat{n} (t) \rangle = \bar{n} \exp{(-\kappa t )} $. To track the photon dynamics, we measure the average qubit contrast within a 10-ns demodulation window. The squared modulus of signal contrast $|S_0-S_1|^2$ is proportional to the average photon number~\cite{ryan_tomography_2015,bultink_general_2018}, where $S_0$ ($S_1$) is the readout signal for qubit at $\ket{0}$ ($\ket{1}$) state. We collect and average 65536 shots at each delay position shown in Fig.~\ref{fig:photontime}. By fitting the photon dynamics $\langle \hat{n} (t) \rangle$ to these data points, we can extract $\kappa$ and calculate the compensation flux pulse shape.

\section{Flux pulse calibration\label{sec:flux_ramsey}}

\begin{figure}
    \centering
    \includegraphics[width=\linewidth]{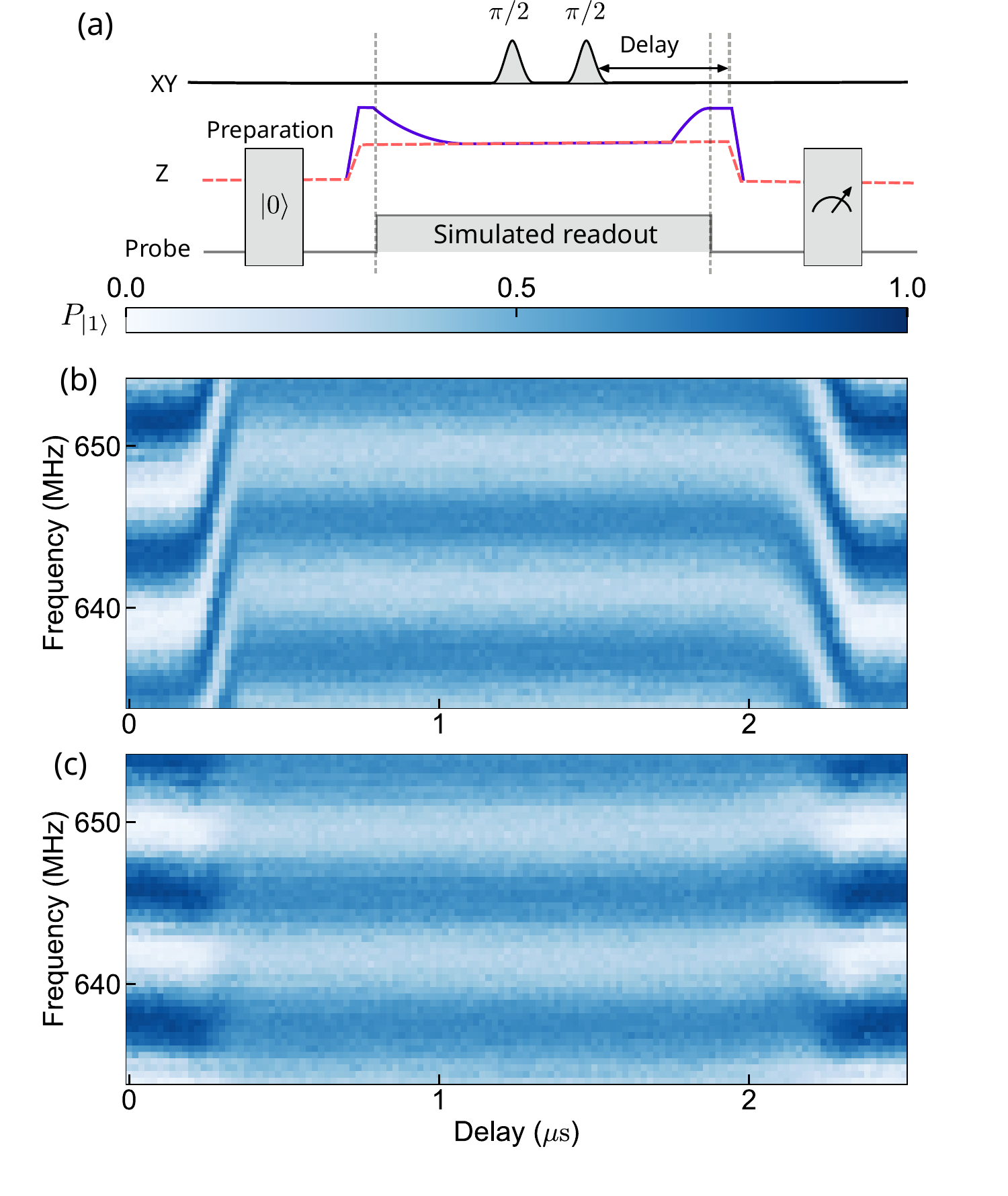}
    \caption{
       Flux pulse calibration. (a) Ramsey sequence to monitor the qubit frequency with (without) the flux compensation, corresponding to the purple (red) line on the Z line. The $\pi/2$ pulse interval is fixed at 100 ns. (b) Ramsey fringes when sweeping qubit pulse frequency and delay between the last $\pi/2$ pulse and the readout pulse without flux compensation. (c) Ramsey fringes after calibrating flux compensation pulse.
    }
    \label{fig:ramseycheck}
\end{figure}

We use a Ramsey experiment to synchronize and fine-tune the flux pulse (Fig.~\ref{fig:ramseycheck}). The interval between the two $\pi/2$ pulses is fixed at 100 ns and we sweep the frequency of the drive, creating Ramsey fringes at each position across the whole range of the simulated readout pulse. Without flux compensation, the Ramsey fringes shift at the beginning and the end of the simulated readout pulse, as shown in Fig.~\ref{fig:ramseycheck}(b). The smaller contrast at the middle of the readout pulse is due to measurement-induced dephasing. We adjust the position and amplitude of the compensation pulse to align the Ramsey fringes across the whole range. 

\section{MIST error metric\label{sec:mist_err}}

To predict the onset of MIST events, we follow the calculations detailed in reference \cite{Nesterov2024}. Specifically, for each flux position, we start by diagnolizing the interaction Hamiltonian (Eq.(\ref{eq:H})). To ensure the correct labeling of the eigenstate $\ket{k,n}$ for near-resonant interactions, where $k$ enumerates the qubit excitation and $n$ is the resonator photon number, we perform the discrete adiabatic state identification (DASI) algorithm: starting from the non-interacting eigenstates $\ket{k,n} \equiv \overline{\ket{k,n}}_{g = 0}$, we incrementally increase the coupling strength by $\delta g = 1~\text{MHz}$ until we reach the target coupling strength of $g_{na} = 53.7~\text{MHz}$. At each new step, we diagonalize the updated Hamiltonian, and the eigenstate $\overline{\ket{k,n}}_{g+\delta g}$ is identified by maximizing its overlaps with the previous eigenstate $\overline{\ket{k,n}}_{g}$. 
Before the onset of MIST, the coupled system during measurement is well described by dressed coherent states, where the eigenstate for an arbitrary coherent amplitude $\alpha$,
\begin{equation}
    \begin{aligned}
        \ket{k,\alpha} = e^{-|\alpha|^2/2}\sum_n \frac{\alpha^2}{\sqrt{n!}}\overline{\ket{k,n}},
    \end{aligned}
\end{equation}
is found from a coherent mixture of the final eigenstates $\overline{\ket{k,n}}$  at the target coupling strength $g_{na}$.
Consequently, the error metric
\begin{equation}
    \begin{aligned}
        \epsilon_\text{MIST} = \left| 1- \frac{\bra{k,\alpha}\hat{a}\ket{k,\alpha}}{\alpha} \right|
    \end{aligned}
\end{equation}
quantifies the extent to which the coherent-dressed state remains a valid description of the driven system with the removal of a single photon. This metric is trivially zero for a coherent state, whereas a large $\epsilon_\text{MIST}$ indicates the onset of MIST. Finally, we relate the coherent amplitude to the readout drive power, or mean resonator photon number $\braket{n}$, according to $\braket{n} = |\alpha|^2$.

\bibliography{references}

\begin{thebibliography}{45}%
\makeatletter
\providecommand \@ifxundefined [1]{%
 \@ifx{#1\undefined}
}%
\providecommand \@ifnum [1]{%
 \ifnum #1\expandafter \@firstoftwo
 \else \expandafter \@secondoftwo
 \fi
}%
\providecommand \@ifx [1]{%
 \ifx #1\expandafter \@firstoftwo
 \else \expandafter \@secondoftwo
 \fi
}%
\providecommand \natexlab [1]{#1}%
\providecommand \enquote  [1]{``#1''}%
\providecommand \bibnamefont  [1]{#1}%
\providecommand \bibfnamefont [1]{#1}%
\providecommand \citenamefont [1]{#1}%
\providecommand \href@noop [0]{\@secondoftwo}%
\providecommand \href [0]{\begingroup \@sanitize@url \@href}%
\providecommand \@href[1]{\@@startlink{#1}\@@href}%
\providecommand \@@href[1]{\endgroup#1\@@endlink}%
\providecommand \@sanitize@url [0]{\catcode `\\12\catcode `\$12\catcode
  `\&12\catcode `\#12\catcode `\^12\catcode `\_12\catcode `\%12\relax}%
\providecommand \@@startlink[1]{}%
\providecommand \@@endlink[0]{}%
\providecommand \url  [0]{\begingroup\@sanitize@url \@url }%
\providecommand \@url [1]{\endgroup\@href {#1}{\urlprefix }}%
\providecommand \urlprefix  [0]{URL }%
\providecommand \Eprint [0]{\href }%
\providecommand \doibase [0]{https://doi.org/}%
\providecommand \selectlanguage [0]{\@gobble}%
\providecommand \bibinfo  [0]{\@secondoftwo}%
\providecommand \bibfield  [0]{\@secondoftwo}%
\providecommand \translation [1]{[#1]}%
\providecommand \BibitemOpen [0]{}%
\providecommand \bibitemStop [0]{}%
\providecommand \bibitemNoStop [0]{.\EOS\space}%
\providecommand \EOS [0]{\spacefactor3000\relax}%
\providecommand \BibitemShut  [1]{\csname bibitem#1\endcsname}%
\let\auto@bib@innerbib\@empty
\bibitem [{\citenamefont {Blais}\ \emph {et~al.}(2021)\citenamefont {Blais},
  \citenamefont {Grimsmo}, \citenamefont {Girvin},\ and\ \citenamefont
  {Wallraff}}]{Blais_review}%
  \BibitemOpen
  \bibfield  {author} {\bibinfo {author} {\bibfnamefont {A.}~\bibnamefont
  {Blais}}, \bibinfo {author} {\bibfnamefont {A.~L.}\ \bibnamefont {Grimsmo}},
  \bibinfo {author} {\bibfnamefont {S.~M.}\ \bibnamefont {Girvin}},\ and\
  \bibinfo {author} {\bibfnamefont {A.}~\bibnamefont {Wallraff}},\ }\bibfield
  {title} {\bibinfo {title} {Circuit quantum electrodynamics},\ }\href
  {https://doi.org/10.1103/RevModPhys.93.025005} {\bibfield  {journal}
  {\bibinfo  {journal} {Rev. Mod. Phys.}\ }\textbf {\bibinfo {volume} {93}},\
  \bibinfo {pages} {025005} (\bibinfo {year} {2021})}\BibitemShut {NoStop}%
\bibitem [{\citenamefont {Sunada}\ \emph {et~al.}(2022)\citenamefont {Sunada},
  \citenamefont {Kono}, \citenamefont {Ilves}, \citenamefont {Tamate},
  \citenamefont {Sugiyama}, \citenamefont {Tabuchi},\ and\ \citenamefont
  {Nakamura}}]{Sunada2022}%
  \BibitemOpen
  \bibfield  {author} {\bibinfo {author} {\bibfnamefont {Y.}~\bibnamefont
  {Sunada}}, \bibinfo {author} {\bibfnamefont {S.}~\bibnamefont {Kono}},
  \bibinfo {author} {\bibfnamefont {J.}~\bibnamefont {Ilves}}, \bibinfo
  {author} {\bibfnamefont {S.}~\bibnamefont {Tamate}}, \bibinfo {author}
  {\bibfnamefont {T.}~\bibnamefont {Sugiyama}}, \bibinfo {author}
  {\bibfnamefont {Y.}~\bibnamefont {Tabuchi}},\ and\ \bibinfo {author}
  {\bibfnamefont {Y.}~\bibnamefont {Nakamura}},\ }\bibfield  {title} {\bibinfo
  {title} {Fast readout and reset of a superconducting qubit coupled to a
  resonator with an intrinsic purcell filter},\ }\href
  {https://doi.org/10.1103/PhysRevApplied.17.044016} {\bibfield  {journal}
  {\bibinfo  {journal} {Phys. Rev. Appl.}\ }\textbf {\bibinfo {volume} {17}},\
  \bibinfo {pages} {044016} (\bibinfo {year} {2022})}\BibitemShut {NoStop}%
\bibitem [{\citenamefont {Sunada}\ \emph {et~al.}(2024)\citenamefont {Sunada},
  \citenamefont {Yuki}, \citenamefont {Wang}, \citenamefont {Miyamura},
  \citenamefont {Ilves}, \citenamefont {Matsuura}, \citenamefont {Spring},
  \citenamefont {Tamate}, \citenamefont {Kono},\ and\ \citenamefont
  {Nakamura}}]{Sunada2024}%
  \BibitemOpen
  \bibfield  {author} {\bibinfo {author} {\bibfnamefont {Y.}~\bibnamefont
  {Sunada}}, \bibinfo {author} {\bibfnamefont {K.}~\bibnamefont {Yuki}},
  \bibinfo {author} {\bibfnamefont {Z.}~\bibnamefont {Wang}}, \bibinfo {author}
  {\bibfnamefont {T.}~\bibnamefont {Miyamura}}, \bibinfo {author}
  {\bibfnamefont {J.}~\bibnamefont {Ilves}}, \bibinfo {author} {\bibfnamefont
  {K.}~\bibnamefont {Matsuura}}, \bibinfo {author} {\bibfnamefont {P.~A.}\
  \bibnamefont {Spring}}, \bibinfo {author} {\bibfnamefont {S.}~\bibnamefont
  {Tamate}}, \bibinfo {author} {\bibfnamefont {S.}~\bibnamefont {Kono}},\ and\
  \bibinfo {author} {\bibfnamefont {Y.}~\bibnamefont {Nakamura}},\ }\bibfield
  {title} {\bibinfo {title} {Photon-noise-tolerant dispersive readout of a
  superconducting qubit using a nonlinear purcell filter},\ }\href
  {https://doi.org/10.1103/PRXQuantum.5.010307} {\bibfield  {journal} {\bibinfo
   {journal} {PRX Quantum}\ }\textbf {\bibinfo {volume} {5}},\ \bibinfo {pages}
  {010307} (\bibinfo {year} {2024})}\BibitemShut {NoStop}%
\bibitem [{\citenamefont {Walter}\ \emph {et~al.}(2017)\citenamefont {Walter},
  \citenamefont {Kurpiers}, \citenamefont {Gasparinetti}, \citenamefont
  {Magnard}, \citenamefont {Poto\ifmmode~\check{c}\else \v{c}\fi{}nik},
  \citenamefont {Salath\'e}, \citenamefont {Pechal}, \citenamefont {Mondal},
  \citenamefont {Oppliger}, \citenamefont {Eichler},\ and\ \citenamefont
  {Wallraff}}]{Walter2017}%
  \BibitemOpen
  \bibfield  {author} {\bibinfo {author} {\bibfnamefont {T.}~\bibnamefont
  {Walter}}, \bibinfo {author} {\bibfnamefont {P.}~\bibnamefont {Kurpiers}},
  \bibinfo {author} {\bibfnamefont {S.}~\bibnamefont {Gasparinetti}}, \bibinfo
  {author} {\bibfnamefont {P.}~\bibnamefont {Magnard}}, \bibinfo {author}
  {\bibfnamefont {A.}~\bibnamefont {Poto\ifmmode~\check{c}\else
  \v{c}\fi{}nik}}, \bibinfo {author} {\bibfnamefont {Y.}~\bibnamefont
  {Salath\'e}}, \bibinfo {author} {\bibfnamefont {M.}~\bibnamefont {Pechal}},
  \bibinfo {author} {\bibfnamefont {M.}~\bibnamefont {Mondal}}, \bibinfo
  {author} {\bibfnamefont {M.}~\bibnamefont {Oppliger}}, \bibinfo {author}
  {\bibfnamefont {C.}~\bibnamefont {Eichler}},\ and\ \bibinfo {author}
  {\bibfnamefont {A.}~\bibnamefont {Wallraff}},\ }\bibfield  {title} {\bibinfo
  {title} {Rapid high-fidelity single-shot dispersive readout of
  superconducting qubits},\ }\href
  {https://doi.org/10.1103/PhysRevApplied.7.054020} {\bibfield  {journal}
  {\bibinfo  {journal} {Phys. Rev. Appl.}\ }\textbf {\bibinfo {volume} {7}},\
  \bibinfo {pages} {054020} (\bibinfo {year} {2017})}\BibitemShut {NoStop}%
\bibitem [{\citenamefont {Swiadek}\ \emph {et~al.}(2023)\citenamefont
  {Swiadek}, \citenamefont {Shillito}, \citenamefont {Magnard}, \citenamefont
  {Remm}, \citenamefont {Hellings}, \citenamefont {Lacroix}, \citenamefont
  {Ficheux}, \citenamefont {Zanuz}, \citenamefont {Norris}, \citenamefont
  {Blais}, \citenamefont {Krinner},\ and\ \citenamefont
  {Wallraff}}]{Swiadek2024}%
  \BibitemOpen
  \bibfield  {author} {\bibinfo {author} {\bibfnamefont {F.}~\bibnamefont
  {Swiadek}}, \bibinfo {author} {\bibfnamefont {R.}~\bibnamefont {Shillito}},
  \bibinfo {author} {\bibfnamefont {P.}~\bibnamefont {Magnard}}, \bibinfo
  {author} {\bibfnamefont {A.}~\bibnamefont {Remm}}, \bibinfo {author}
  {\bibfnamefont {C.}~\bibnamefont {Hellings}}, \bibinfo {author}
  {\bibfnamefont {N.}~\bibnamefont {Lacroix}}, \bibinfo {author} {\bibfnamefont
  {Q.}~\bibnamefont {Ficheux}}, \bibinfo {author} {\bibfnamefont {D.~C.}\
  \bibnamefont {Zanuz}}, \bibinfo {author} {\bibfnamefont {G.~J.}\ \bibnamefont
  {Norris}}, \bibinfo {author} {\bibfnamefont {A.}~\bibnamefont {Blais}},
  \bibinfo {author} {\bibfnamefont {S.}~\bibnamefont {Krinner}},\ and\ \bibinfo
  {author} {\bibfnamefont {A.}~\bibnamefont {Wallraff}},\ }\href
  {http://arxiv.org/abs/2307.07765} {\bibinfo {title} {Enhancing {Dispersive}
  {Readout} of {Superconducting} {Qubits} {Through} {Dynamic} {Control} of the
  {Dispersive} {Shift}: {Experiment} and {Theory}}} (\bibinfo {year} {2023}),\
  \bibinfo {note} {arXiv:2307.07765 [quant-ph]}\BibitemShut {NoStop}%
\bibitem [{\citenamefont {Stefanski}\ \emph {et~al.}(2024)\citenamefont
  {Stefanski}, \citenamefont {Yilmaz}, \citenamefont {Huang}, \citenamefont
  {Zwanenburg}, \citenamefont {Singh}, \citenamefont {Wang}, \citenamefont
  {Splitthoff},\ and\ \citenamefont {Andersen}}]{stefanski_improved_2024}%
  \BibitemOpen
  \bibfield  {author} {\bibinfo {author} {\bibfnamefont {T.~V.}\ \bibnamefont
  {Stefanski}}, \bibinfo {author} {\bibfnamefont {F.}~\bibnamefont {Yilmaz}},
  \bibinfo {author} {\bibfnamefont {E.~Y.}\ \bibnamefont {Huang}}, \bibinfo
  {author} {\bibfnamefont {M.~F.~S.}\ \bibnamefont {Zwanenburg}}, \bibinfo
  {author} {\bibfnamefont {S.}~\bibnamefont {Singh}}, \bibinfo {author}
  {\bibfnamefont {S.}~\bibnamefont {Wang}}, \bibinfo {author} {\bibfnamefont
  {L.~J.}\ \bibnamefont {Splitthoff}},\ and\ \bibinfo {author} {\bibfnamefont
  {C.~K.}\ \bibnamefont {Andersen}},\ }\href
  {https://doi.org/10.48550/arXiv.2411.13437} {\bibinfo {title} {Improved
  fluxonium readout through dynamic flux pulsing}} (\bibinfo {year} {2024}),\
  \bibinfo {note} {arXiv:2411.13437 [quant-ph]}\BibitemShut {NoStop}%
\bibitem [{\citenamefont {Bothara}\ \emph {et~al.}(2025)\citenamefont
  {Bothara}, \citenamefont {Das}, \citenamefont {Salunkhe}, \citenamefont
  {Chand}, \citenamefont {Deshmukh}, \citenamefont {Patankar},\ and\
  \citenamefont {Vijay}}]{Bothara2025}%
  \BibitemOpen
  \bibfield  {author} {\bibinfo {author} {\bibfnamefont {G.}~\bibnamefont
  {Bothara}}, \bibinfo {author} {\bibfnamefont {S.}~\bibnamefont {Das}},
  \bibinfo {author} {\bibfnamefont {K.~V.}\ \bibnamefont {Salunkhe}}, \bibinfo
  {author} {\bibfnamefont {M.}~\bibnamefont {Chand}}, \bibinfo {author}
  {\bibfnamefont {J.}~\bibnamefont {Deshmukh}}, \bibinfo {author}
  {\bibfnamefont {M.~P.}\ \bibnamefont {Patankar}},\ and\ \bibinfo {author}
  {\bibfnamefont {R.}~\bibnamefont {Vijay}},\ }\href
  {https://doi.org/10.48550/arXiv.2501.16691} {\bibinfo {title} {High-fidelity
  {QND} readout and measurement back-action in a {Tantalum}-based
  high-coherence fluxonium qubit}} (\bibinfo {year} {2025}),\ \bibinfo {note}
  {arXiv:2501.16691 [quant-ph]}\BibitemShut {NoStop}%
\bibitem [{\citenamefont {Levine}\ \emph {et~al.}(2024)\citenamefont {Levine},
  \citenamefont {Haim}, \citenamefont {Hung}, \citenamefont {Alidoust},
  \citenamefont {Kalaee}, \citenamefont {DeLorenzo}, \citenamefont {Wollack},
  \citenamefont {Arrangoiz-Arriola}, \citenamefont {Khalajhedayati},
  \citenamefont {Sanil}, \citenamefont {Moradinejad}, \citenamefont {Vaknin},
  \citenamefont {Kubica}, \citenamefont {Hover}, \citenamefont {Aghaeimeibodi},
  \citenamefont {Alcid}, \citenamefont {Baek}, \citenamefont {Barnett},
  \citenamefont {Bawdekar}, \citenamefont {Bienias}, \citenamefont {Carson},
  \citenamefont {Chen}, \citenamefont {Chen}, \citenamefont {Chinkezian},
  \citenamefont {Chisholm}, \citenamefont {Clifford}, \citenamefont {Cosmic},
  \citenamefont {Crisosto}, \citenamefont {Dalzell}, \citenamefont {Davis},
  \citenamefont {D'Ewart}, \citenamefont {Diez}, \citenamefont {D'Souza},
  \citenamefont {Dumitrescu}, \citenamefont {Elkhouly}, \citenamefont {Fang},
  \citenamefont {Fang}, \citenamefont {Flammia}, \citenamefont {Fling},
  \citenamefont {Garcia}, \citenamefont {Gharzai}, \citenamefont {Gorshkov},
  \citenamefont {Gray}, \citenamefont {Grimberg}, \citenamefont {Grimsmo},
  \citenamefont {Hann}, \citenamefont {He}, \citenamefont {Heidel},
  \citenamefont {Howell}, \citenamefont {Hunt}, \citenamefont {Iverson},
  \citenamefont {Jarrige}, \citenamefont {Jiang}, \citenamefont {Jones},
  \citenamefont {Karabalin}, \citenamefont {Karalekas}, \citenamefont {Keller},
  \citenamefont {Lasi}, \citenamefont {Lee}, \citenamefont {Ly}, \citenamefont
  {MacCabe}, \citenamefont {Mahuli}, \citenamefont {Marcaud}, \citenamefont
  {Matheny}, \citenamefont {McArdle}, \citenamefont {McCabe}, \citenamefont
  {Merton}, \citenamefont {Miles}, \citenamefont {Milsted}, \citenamefont
  {Mishra}, \citenamefont {Moncelsi}, \citenamefont {Naghiloo}, \citenamefont
  {Noh}, \citenamefont {Oblepias}, \citenamefont {Ortuno}, \citenamefont
  {Owens}, \citenamefont {Pagdilao}, \citenamefont {Panduro}, \citenamefont
  {Paquette}, \citenamefont {Patel}, \citenamefont {Peairs}, \citenamefont
  {Perello}, \citenamefont {Peterson}, \citenamefont {Ponte}, \citenamefont
  {Putterman}, \citenamefont {Refael}, \citenamefont {Reinhold}, \citenamefont
  {Resnick}, \citenamefont {Reyna}, \citenamefont {Rodriguez}, \citenamefont
  {Rose}, \citenamefont {Rubin}, \citenamefont {Runyan}, \citenamefont {Ryan},
  \citenamefont {Sahmoud}, \citenamefont {Scaffidi}, \citenamefont {Shah},
  \citenamefont {Siavoshi}, \citenamefont {Sivarajah}, \citenamefont
  {Skogland}, \citenamefont {Su}, \citenamefont {Swenson}, \citenamefont
  {Sylvia}, \citenamefont {Teo}, \citenamefont {Tomada}, \citenamefont
  {Torlai}, \citenamefont {Wistrom}, \citenamefont {Zhang}, \citenamefont
  {Zuk}, \citenamefont {Clerk}, \citenamefont {Brand\~ao}, \citenamefont
  {Retzker},\ and\ \citenamefont {Painter}}]{Levine2024}%
  \BibitemOpen
  \bibfield  {author} {\bibinfo {author} {\bibfnamefont {H.}~\bibnamefont
  {Levine}}, \bibinfo {author} {\bibfnamefont {A.}~\bibnamefont {Haim}},
  \bibinfo {author} {\bibfnamefont {J.~S.~C.}\ \bibnamefont {Hung}}, \bibinfo
  {author} {\bibfnamefont {N.}~\bibnamefont {Alidoust}}, \bibinfo {author}
  {\bibfnamefont {M.}~\bibnamefont {Kalaee}}, \bibinfo {author} {\bibfnamefont
  {L.}~\bibnamefont {DeLorenzo}}, \bibinfo {author} {\bibfnamefont {E.~A.}\
  \bibnamefont {Wollack}}, \bibinfo {author} {\bibfnamefont {P.}~\bibnamefont
  {Arrangoiz-Arriola}}, \bibinfo {author} {\bibfnamefont {A.}~\bibnamefont
  {Khalajhedayati}}, \bibinfo {author} {\bibfnamefont {R.}~\bibnamefont
  {Sanil}}, \bibinfo {author} {\bibfnamefont {H.}~\bibnamefont {Moradinejad}},
  \bibinfo {author} {\bibfnamefont {Y.}~\bibnamefont {Vaknin}}, \bibinfo
  {author} {\bibfnamefont {A.}~\bibnamefont {Kubica}}, \bibinfo {author}
  {\bibfnamefont {D.}~\bibnamefont {Hover}}, \bibinfo {author} {\bibfnamefont
  {S.}~\bibnamefont {Aghaeimeibodi}}, \bibinfo {author} {\bibfnamefont {J.~A.}\
  \bibnamefont {Alcid}}, \bibinfo {author} {\bibfnamefont {C.}~\bibnamefont
  {Baek}}, \bibinfo {author} {\bibfnamefont {J.}~\bibnamefont {Barnett}},
  \bibinfo {author} {\bibfnamefont {K.}~\bibnamefont {Bawdekar}}, \bibinfo
  {author} {\bibfnamefont {P.}~\bibnamefont {Bienias}}, \bibinfo {author}
  {\bibfnamefont {H.~A.}\ \bibnamefont {Carson}}, \bibinfo {author}
  {\bibfnamefont {C.}~\bibnamefont {Chen}}, \bibinfo {author} {\bibfnamefont
  {L.}~\bibnamefont {Chen}}, \bibinfo {author} {\bibfnamefont {H.}~\bibnamefont
  {Chinkezian}}, \bibinfo {author} {\bibfnamefont {E.~M.}\ \bibnamefont
  {Chisholm}}, \bibinfo {author} {\bibfnamefont {A.}~\bibnamefont {Clifford}},
  \bibinfo {author} {\bibfnamefont {R.}~\bibnamefont {Cosmic}}, \bibinfo
  {author} {\bibfnamefont {N.}~\bibnamefont {Crisosto}}, \bibinfo {author}
  {\bibfnamefont {A.~M.}\ \bibnamefont {Dalzell}}, \bibinfo {author}
  {\bibfnamefont {E.}~\bibnamefont {Davis}}, \bibinfo {author} {\bibfnamefont
  {J.~M.}\ \bibnamefont {D'Ewart}}, \bibinfo {author} {\bibfnamefont
  {S.}~\bibnamefont {Diez}}, \bibinfo {author} {\bibfnamefont {N.}~\bibnamefont
  {D'Souza}}, \bibinfo {author} {\bibfnamefont {P.~T.}\ \bibnamefont
  {Dumitrescu}}, \bibinfo {author} {\bibfnamefont {E.}~\bibnamefont
  {Elkhouly}}, \bibinfo {author} {\bibfnamefont {M.~T.}\ \bibnamefont {Fang}},
  \bibinfo {author} {\bibfnamefont {Y.}~\bibnamefont {Fang}}, \bibinfo {author}
  {\bibfnamefont {S.}~\bibnamefont {Flammia}}, \bibinfo {author} {\bibfnamefont
  {M.~J.}\ \bibnamefont {Fling}}, \bibinfo {author} {\bibfnamefont
  {G.}~\bibnamefont {Garcia}}, \bibinfo {author} {\bibfnamefont {M.~K.}\
  \bibnamefont {Gharzai}}, \bibinfo {author} {\bibfnamefont {A.~V.}\
  \bibnamefont {Gorshkov}}, \bibinfo {author} {\bibfnamefont {M.~J.}\
  \bibnamefont {Gray}}, \bibinfo {author} {\bibfnamefont {S.}~\bibnamefont
  {Grimberg}}, \bibinfo {author} {\bibfnamefont {A.~L.}\ \bibnamefont
  {Grimsmo}}, \bibinfo {author} {\bibfnamefont {C.~T.}\ \bibnamefont {Hann}},
  \bibinfo {author} {\bibfnamefont {Y.}~\bibnamefont {He}}, \bibinfo {author}
  {\bibfnamefont {S.}~\bibnamefont {Heidel}}, \bibinfo {author} {\bibfnamefont
  {S.}~\bibnamefont {Howell}}, \bibinfo {author} {\bibfnamefont
  {M.}~\bibnamefont {Hunt}}, \bibinfo {author} {\bibfnamefont {J.}~\bibnamefont
  {Iverson}}, \bibinfo {author} {\bibfnamefont {I.}~\bibnamefont {Jarrige}},
  \bibinfo {author} {\bibfnamefont {L.}~\bibnamefont {Jiang}}, \bibinfo
  {author} {\bibfnamefont {W.~M.}\ \bibnamefont {Jones}}, \bibinfo {author}
  {\bibfnamefont {R.}~\bibnamefont {Karabalin}}, \bibinfo {author}
  {\bibfnamefont {P.~J.}\ \bibnamefont {Karalekas}}, \bibinfo {author}
  {\bibfnamefont {A.~J.}\ \bibnamefont {Keller}}, \bibinfo {author}
  {\bibfnamefont {D.}~\bibnamefont {Lasi}}, \bibinfo {author} {\bibfnamefont
  {M.}~\bibnamefont {Lee}}, \bibinfo {author} {\bibfnamefont {V.}~\bibnamefont
  {Ly}}, \bibinfo {author} {\bibfnamefont {G.}~\bibnamefont {MacCabe}},
  \bibinfo {author} {\bibfnamefont {N.}~\bibnamefont {Mahuli}}, \bibinfo
  {author} {\bibfnamefont {G.}~\bibnamefont {Marcaud}}, \bibinfo {author}
  {\bibfnamefont {M.~H.}\ \bibnamefont {Matheny}}, \bibinfo {author}
  {\bibfnamefont {S.}~\bibnamefont {McArdle}}, \bibinfo {author} {\bibfnamefont
  {G.}~\bibnamefont {McCabe}}, \bibinfo {author} {\bibfnamefont
  {G.}~\bibnamefont {Merton}}, \bibinfo {author} {\bibfnamefont
  {C.}~\bibnamefont {Miles}}, \bibinfo {author} {\bibfnamefont
  {A.}~\bibnamefont {Milsted}}, \bibinfo {author} {\bibfnamefont
  {A.}~\bibnamefont {Mishra}}, \bibinfo {author} {\bibfnamefont
  {L.}~\bibnamefont {Moncelsi}}, \bibinfo {author} {\bibfnamefont
  {M.}~\bibnamefont {Naghiloo}}, \bibinfo {author} {\bibfnamefont
  {K.}~\bibnamefont {Noh}}, \bibinfo {author} {\bibfnamefont {E.}~\bibnamefont
  {Oblepias}}, \bibinfo {author} {\bibfnamefont {G.}~\bibnamefont {Ortuno}},
  \bibinfo {author} {\bibfnamefont {J.~C.}\ \bibnamefont {Owens}}, \bibinfo
  {author} {\bibfnamefont {J.}~\bibnamefont {Pagdilao}}, \bibinfo {author}
  {\bibfnamefont {A.}~\bibnamefont {Panduro}}, \bibinfo {author} {\bibfnamefont
  {J.-P.}\ \bibnamefont {Paquette}}, \bibinfo {author} {\bibfnamefont {R.~N.}\
  \bibnamefont {Patel}}, \bibinfo {author} {\bibfnamefont {G.}~\bibnamefont
  {Peairs}}, \bibinfo {author} {\bibfnamefont {D.~J.}\ \bibnamefont {Perello}},
  \bibinfo {author} {\bibfnamefont {E.~C.}\ \bibnamefont {Peterson}}, \bibinfo
  {author} {\bibfnamefont {S.}~\bibnamefont {Ponte}}, \bibinfo {author}
  {\bibfnamefont {H.}~\bibnamefont {Putterman}}, \bibinfo {author}
  {\bibfnamefont {G.}~\bibnamefont {Refael}}, \bibinfo {author} {\bibfnamefont
  {P.}~\bibnamefont {Reinhold}}, \bibinfo {author} {\bibfnamefont
  {R.}~\bibnamefont {Resnick}}, \bibinfo {author} {\bibfnamefont {O.~A.}\
  \bibnamefont {Reyna}}, \bibinfo {author} {\bibfnamefont {R.}~\bibnamefont
  {Rodriguez}}, \bibinfo {author} {\bibfnamefont {J.}~\bibnamefont {Rose}},
  \bibinfo {author} {\bibfnamefont {A.~H.}\ \bibnamefont {Rubin}}, \bibinfo
  {author} {\bibfnamefont {M.}~\bibnamefont {Runyan}}, \bibinfo {author}
  {\bibfnamefont {C.~A.}\ \bibnamefont {Ryan}}, \bibinfo {author}
  {\bibfnamefont {A.}~\bibnamefont {Sahmoud}}, \bibinfo {author} {\bibfnamefont
  {T.}~\bibnamefont {Scaffidi}}, \bibinfo {author} {\bibfnamefont
  {B.}~\bibnamefont {Shah}}, \bibinfo {author} {\bibfnamefont {S.}~\bibnamefont
  {Siavoshi}}, \bibinfo {author} {\bibfnamefont {P.}~\bibnamefont {Sivarajah}},
  \bibinfo {author} {\bibfnamefont {T.}~\bibnamefont {Skogland}}, \bibinfo
  {author} {\bibfnamefont {C.-J.}\ \bibnamefont {Su}}, \bibinfo {author}
  {\bibfnamefont {L.~J.}\ \bibnamefont {Swenson}}, \bibinfo {author}
  {\bibfnamefont {J.}~\bibnamefont {Sylvia}}, \bibinfo {author} {\bibfnamefont
  {S.~M.}\ \bibnamefont {Teo}}, \bibinfo {author} {\bibfnamefont
  {A.}~\bibnamefont {Tomada}}, \bibinfo {author} {\bibfnamefont
  {G.}~\bibnamefont {Torlai}}, \bibinfo {author} {\bibfnamefont
  {M.}~\bibnamefont {Wistrom}}, \bibinfo {author} {\bibfnamefont
  {K.}~\bibnamefont {Zhang}}, \bibinfo {author} {\bibfnamefont
  {I.}~\bibnamefont {Zuk}}, \bibinfo {author} {\bibfnamefont {A.~A.}\
  \bibnamefont {Clerk}}, \bibinfo {author} {\bibfnamefont {F.~G. S.~L.}\
  \bibnamefont {Brand\~ao}}, \bibinfo {author} {\bibfnamefont {A.}~\bibnamefont
  {Retzker}},\ and\ \bibinfo {author} {\bibfnamefont {O.}~\bibnamefont
  {Painter}},\ }\bibfield  {title} {\bibinfo {title} {Demonstrating a
  long-coherence dual-rail erasure qubit using tunable transmons},\ }\href
  {https://doi.org/10.1103/PhysRevX.14.011051} {\bibfield  {journal} {\bibinfo
  {journal} {Phys. Rev. X}\ }\textbf {\bibinfo {volume} {14}},\ \bibinfo
  {pages} {011051} (\bibinfo {year} {2024})}\BibitemShut {NoStop}%
\bibitem [{\citenamefont {Koottandavida}\ \emph {et~al.}(2024)\citenamefont
  {Koottandavida}, \citenamefont {Tsioutsios}, \citenamefont {Kargioti},
  \citenamefont {Smith}, \citenamefont {Joshi}, \citenamefont {Dai},
  \citenamefont {Teoh}, \citenamefont {Curtis}, \citenamefont {Frunzio},
  \citenamefont {Schoelkopf},\ and\ \citenamefont
  {Devoret}}]{Koottandavida2024}%
  \BibitemOpen
  \bibfield  {author} {\bibinfo {author} {\bibfnamefont {A.}~\bibnamefont
  {Koottandavida}}, \bibinfo {author} {\bibfnamefont {I.}~\bibnamefont
  {Tsioutsios}}, \bibinfo {author} {\bibfnamefont {A.}~\bibnamefont
  {Kargioti}}, \bibinfo {author} {\bibfnamefont {C.~R.}\ \bibnamefont {Smith}},
  \bibinfo {author} {\bibfnamefont {V.~R.}\ \bibnamefont {Joshi}}, \bibinfo
  {author} {\bibfnamefont {W.}~\bibnamefont {Dai}}, \bibinfo {author}
  {\bibfnamefont {J.~D.}\ \bibnamefont {Teoh}}, \bibinfo {author}
  {\bibfnamefont {J.~C.}\ \bibnamefont {Curtis}}, \bibinfo {author}
  {\bibfnamefont {L.}~\bibnamefont {Frunzio}}, \bibinfo {author} {\bibfnamefont
  {R.~J.}\ \bibnamefont {Schoelkopf}},\ and\ \bibinfo {author} {\bibfnamefont
  {M.~H.}\ \bibnamefont {Devoret}},\ }\bibfield  {title} {\bibinfo {title}
  {Erasure detection of a dual-rail qubit encoded in a double-post
  superconducting cavity},\ }\href
  {https://doi.org/10.1103/PhysRevLett.132.180601} {\bibfield  {journal}
  {\bibinfo  {journal} {Phys. Rev. Lett.}\ }\textbf {\bibinfo {volume} {132}},\
  \bibinfo {pages} {180601} (\bibinfo {year} {2024})}\BibitemShut {NoStop}%
\bibitem [{\citenamefont {{Google Quantum AI}}(2023)}]{Google2023}%
  \BibitemOpen
  \bibfield  {author} {\bibinfo {author} {\bibnamefont {{Google Quantum AI}}},\
  }\bibfield  {title} {\bibinfo {title} {Suppressing quantum errors by scaling
  a surface code logical qubit},\ }\href
  {https://doi.org/10.1038/s41586-022-05434-1} {\bibfield  {journal} {\bibinfo
  {journal} {Nature}\ }\textbf {\bibinfo {volume} {614}},\ \bibinfo {pages}
  {676} (\bibinfo {year} {2023})},\ \bibinfo {note} {published online: 22
  February 2023}\BibitemShut {NoStop}%
\bibitem [{\citenamefont {Ni}\ \emph {et~al.}(2023)\citenamefont {Ni},
  \citenamefont {Li}, \citenamefont {Deng}, \citenamefont {Cai}, \citenamefont
  {Wang}, \citenamefont {Yang}, \citenamefont {Liu}, \citenamefont {Ivanov},
  \citenamefont {Wang}, \citenamefont {Song}, \citenamefont {Zhang},
  \citenamefont {Zhao}, \citenamefont {Liu}, \citenamefont {Chen},
  \citenamefont {Huang}, \citenamefont {Zhong}, \citenamefont {Liu},
  \citenamefont {Zhao}, \citenamefont {Liu}, \citenamefont {Ho}, \citenamefont
  {Benjamin}, \citenamefont {Kim}, \citenamefont {Zhang}, \citenamefont {Li},
  \citenamefont {Monz}, \citenamefont {Wang}, \citenamefont {Chen},
  \citenamefont {Lu},\ and\ \citenamefont {Pan}}]{Ni2023}%
  \BibitemOpen
  \bibfield  {author} {\bibinfo {author} {\bibfnamefont {Z.}~\bibnamefont
  {Ni}}, \bibinfo {author} {\bibfnamefont {K.}~\bibnamefont {Li}}, \bibinfo
  {author} {\bibfnamefont {X.}~\bibnamefont {Deng}}, \bibinfo {author}
  {\bibfnamefont {W.}~\bibnamefont {Cai}}, \bibinfo {author} {\bibfnamefont
  {Y.}~\bibnamefont {Wang}}, \bibinfo {author} {\bibfnamefont {H.}~\bibnamefont
  {Yang}}, \bibinfo {author} {\bibfnamefont {J.}~\bibnamefont {Liu}}, \bibinfo
  {author} {\bibfnamefont {D.}~\bibnamefont {Ivanov}}, \bibinfo {author}
  {\bibfnamefont {X.}~\bibnamefont {Wang}}, \bibinfo {author} {\bibfnamefont
  {F.}~\bibnamefont {Song}}, \bibinfo {author} {\bibfnamefont {P.}~\bibnamefont
  {Zhang}}, \bibinfo {author} {\bibfnamefont {H.}~\bibnamefont {Zhao}},
  \bibinfo {author} {\bibfnamefont {W.}~\bibnamefont {Liu}}, \bibinfo {author}
  {\bibfnamefont {Y.}~\bibnamefont {Chen}}, \bibinfo {author} {\bibfnamefont
  {T.}~\bibnamefont {Huang}}, \bibinfo {author} {\bibfnamefont
  {Y.}~\bibnamefont {Zhong}}, \bibinfo {author} {\bibfnamefont
  {H.}~\bibnamefont {Liu}}, \bibinfo {author} {\bibfnamefont {S.}~\bibnamefont
  {Zhao}}, \bibinfo {author} {\bibfnamefont {S.-Y.}\ \bibnamefont {Liu}},
  \bibinfo {author} {\bibfnamefont {A.}~\bibnamefont {Ho}}, \bibinfo {author}
  {\bibfnamefont {S.}~\bibnamefont {Benjamin}}, \bibinfo {author}
  {\bibfnamefont {K.}~\bibnamefont {Kim}}, \bibinfo {author} {\bibfnamefont
  {X.}~\bibnamefont {Zhang}}, \bibinfo {author} {\bibfnamefont
  {T.}~\bibnamefont {Li}}, \bibinfo {author} {\bibfnamefont {T.}~\bibnamefont
  {Monz}}, \bibinfo {author} {\bibfnamefont {P.}~\bibnamefont {Wang}}, \bibinfo
  {author} {\bibfnamefont {Y.-A.}\ \bibnamefont {Chen}}, \bibinfo {author}
  {\bibfnamefont {C.-Y.}\ \bibnamefont {Lu}},\ and\ \bibinfo {author}
  {\bibfnamefont {J.-W.}\ \bibnamefont {Pan}},\ }\bibfield  {title} {\bibinfo
  {title} {Beating the break-even point with a discrete-variable-encoded
  logical qubit},\ }\href {https://doi.org/10.1038/s41586-023-05784-4}
  {\bibfield  {journal} {\bibinfo  {journal} {Nature}\ }\textbf {\bibinfo
  {volume} {616}},\ \bibinfo {pages} {56} (\bibinfo {year} {2023})},\ \bibinfo
  {note} {published: 05 April 2023}\BibitemShut {NoStop}%
\bibitem [{\citenamefont {Sank}\ \emph {et~al.}(2016)\citenamefont {Sank},
  \citenamefont {Chen}, \citenamefont {Khezri}, \citenamefont {Kelly},
  \citenamefont {Barends}, \citenamefont {Campbell}, \citenamefont {Chen},
  \citenamefont {Chiaro}, \citenamefont {Dunsworth}, \citenamefont {Fowler},
  \citenamefont {Jeffrey}, \citenamefont {Lucero}, \citenamefont {Megrant},
  \citenamefont {Mutus}, \citenamefont {Neeley}, \citenamefont {Neill},
  \citenamefont {O'Malley}, \citenamefont {Quintana}, \citenamefont {Roushan},
  \citenamefont {Vainsencher}, \citenamefont {White}, \citenamefont {Wenner},
  \citenamefont {Korotkov},\ and\ \citenamefont {Martinis}}]{Sank2016}%
  \BibitemOpen
  \bibfield  {author} {\bibinfo {author} {\bibfnamefont {D.}~\bibnamefont
  {Sank}}, \bibinfo {author} {\bibfnamefont {Z.}~\bibnamefont {Chen}}, \bibinfo
  {author} {\bibfnamefont {M.}~\bibnamefont {Khezri}}, \bibinfo {author}
  {\bibfnamefont {J.}~\bibnamefont {Kelly}}, \bibinfo {author} {\bibfnamefont
  {R.}~\bibnamefont {Barends}}, \bibinfo {author} {\bibfnamefont
  {B.}~\bibnamefont {Campbell}}, \bibinfo {author} {\bibfnamefont
  {Y.}~\bibnamefont {Chen}}, \bibinfo {author} {\bibfnamefont {B.}~\bibnamefont
  {Chiaro}}, \bibinfo {author} {\bibfnamefont {A.}~\bibnamefont {Dunsworth}},
  \bibinfo {author} {\bibfnamefont {A.}~\bibnamefont {Fowler}}, \bibinfo
  {author} {\bibfnamefont {E.}~\bibnamefont {Jeffrey}}, \bibinfo {author}
  {\bibfnamefont {E.}~\bibnamefont {Lucero}}, \bibinfo {author} {\bibfnamefont
  {A.}~\bibnamefont {Megrant}}, \bibinfo {author} {\bibfnamefont
  {J.}~\bibnamefont {Mutus}}, \bibinfo {author} {\bibfnamefont
  {M.}~\bibnamefont {Neeley}}, \bibinfo {author} {\bibfnamefont
  {C.}~\bibnamefont {Neill}}, \bibinfo {author} {\bibfnamefont {P.~J.~J.}\
  \bibnamefont {O'Malley}}, \bibinfo {author} {\bibfnamefont {C.}~\bibnamefont
  {Quintana}}, \bibinfo {author} {\bibfnamefont {P.}~\bibnamefont {Roushan}},
  \bibinfo {author} {\bibfnamefont {A.}~\bibnamefont {Vainsencher}}, \bibinfo
  {author} {\bibfnamefont {T.}~\bibnamefont {White}}, \bibinfo {author}
  {\bibfnamefont {J.}~\bibnamefont {Wenner}}, \bibinfo {author} {\bibfnamefont
  {A.~N.}\ \bibnamefont {Korotkov}},\ and\ \bibinfo {author} {\bibfnamefont
  {J.~M.}\ \bibnamefont {Martinis}},\ }\bibfield  {title} {\bibinfo {title}
  {Measurement-induced state transitions in a superconducting qubit: Beyond the
  rotating wave approximation},\ }\href
  {https://doi.org/10.1103/PhysRevLett.117.190503} {\bibfield  {journal}
  {\bibinfo  {journal} {Phys. Rev. Lett.}\ }\textbf {\bibinfo {volume} {117}},\
  \bibinfo {pages} {190503} (\bibinfo {year} {2016})}\BibitemShut {NoStop}%
\bibitem [{\citenamefont {Khezri}\ \emph {et~al.}(2023)\citenamefont {Khezri},
  \citenamefont {Opremcak}, \citenamefont {Chen}, \citenamefont {Miao},
  \citenamefont {McEwen}, \citenamefont {Bengtsson}, \citenamefont {White},
  \citenamefont {Naaman}, \citenamefont {Sank}, \citenamefont {Korotkov},
  \citenamefont {Chen},\ and\ \citenamefont {Smelyanskiy}}]{Khezri2023}%
  \BibitemOpen
  \bibfield  {author} {\bibinfo {author} {\bibfnamefont {M.}~\bibnamefont
  {Khezri}}, \bibinfo {author} {\bibfnamefont {A.}~\bibnamefont {Opremcak}},
  \bibinfo {author} {\bibfnamefont {Z.}~\bibnamefont {Chen}}, \bibinfo {author}
  {\bibfnamefont {K.~C.}\ \bibnamefont {Miao}}, \bibinfo {author}
  {\bibfnamefont {M.}~\bibnamefont {McEwen}}, \bibinfo {author} {\bibfnamefont
  {A.}~\bibnamefont {Bengtsson}}, \bibinfo {author} {\bibfnamefont
  {T.}~\bibnamefont {White}}, \bibinfo {author} {\bibfnamefont
  {O.}~\bibnamefont {Naaman}}, \bibinfo {author} {\bibfnamefont
  {D.}~\bibnamefont {Sank}}, \bibinfo {author} {\bibfnamefont {A.~N.}\
  \bibnamefont {Korotkov}}, \bibinfo {author} {\bibfnamefont {Y.}~\bibnamefont
  {Chen}},\ and\ \bibinfo {author} {\bibfnamefont {V.}~\bibnamefont
  {Smelyanskiy}},\ }\bibfield  {title} {\bibinfo {title} {Measurement-induced
  state transitions in a superconducting qubit: Within the rotating-wave
  approximation},\ }\href {https://doi.org/10.1103/PhysRevApplied.20.054008}
  {\bibfield  {journal} {\bibinfo  {journal} {Phys. Rev. Appl.}\ }\textbf
  {\bibinfo {volume} {20}},\ \bibinfo {pages} {054008} (\bibinfo {year}
  {2023})}\BibitemShut {NoStop}%
\bibitem [{\citenamefont {Bista}\ \emph {et~al.}(2025)\citenamefont {Bista},
  \citenamefont {Thibodeau}, \citenamefont {Nie}, \citenamefont {Chow},
  \citenamefont {Clark},\ and\ \citenamefont {Kou}}]{Bista2025}%
  \BibitemOpen
  \bibfield  {author} {\bibinfo {author} {\bibfnamefont {A.}~\bibnamefont
  {Bista}}, \bibinfo {author} {\bibfnamefont {M.}~\bibnamefont {Thibodeau}},
  \bibinfo {author} {\bibfnamefont {K.}~\bibnamefont {Nie}}, \bibinfo {author}
  {\bibfnamefont {K.}~\bibnamefont {Chow}}, \bibinfo {author} {\bibfnamefont
  {B.~K.}\ \bibnamefont {Clark}},\ and\ \bibinfo {author} {\bibfnamefont
  {A.}~\bibnamefont {Kou}},\ }\href {https://doi.org/10.48550/arXiv.2501.17807}
  {\bibinfo {title} {Readout-induced leakage of the fluxonium qubit}} (\bibinfo
  {year} {2025}),\ \bibinfo {note} {arXiv:2501.17807 [quant-ph]}\BibitemShut
  {NoStop}%
\bibitem [{\citenamefont {Hazra}\ \emph {et~al.}(2025)\citenamefont {Hazra},
  \citenamefont {Dai}, \citenamefont {Connolly}, \citenamefont {Kurilovich},
  \citenamefont {Wang}, \citenamefont {Frunzio},\ and\ \citenamefont
  {Devoret}}]{Hazra2025}%
  \BibitemOpen
  \bibfield  {author} {\bibinfo {author} {\bibfnamefont {S.}~\bibnamefont
  {Hazra}}, \bibinfo {author} {\bibfnamefont {W.}~\bibnamefont {Dai}}, \bibinfo
  {author} {\bibfnamefont {T.}~\bibnamefont {Connolly}}, \bibinfo {author}
  {\bibfnamefont {P.~D.}\ \bibnamefont {Kurilovich}}, \bibinfo {author}
  {\bibfnamefont {Z.}~\bibnamefont {Wang}}, \bibinfo {author} {\bibfnamefont
  {L.}~\bibnamefont {Frunzio}},\ and\ \bibinfo {author} {\bibfnamefont {M.~H.}\
  \bibnamefont {Devoret}},\ }\bibfield  {title} {\bibinfo {title} {Benchmarking
  the readout of a superconducting qubit for repeated measurements},\ }\href
  {https://doi.org/10.1103/PhysRevLett.134.100601} {\bibfield  {journal}
  {\bibinfo  {journal} {Phys. Rev. Lett.}\ }\textbf {\bibinfo {volume} {134}},\
  \bibinfo {pages} {100601} (\bibinfo {year} {2025})}\BibitemShut {NoStop}%
\bibitem [{\citenamefont {Wang}\ \emph {et~al.}(2025)\citenamefont {Wang},
  \citenamefont {D'Anjou}, \citenamefont {Gigon}, \citenamefont {Blais},\ and\
  \citenamefont {Blok}}]{Wang2025}%
  \BibitemOpen
  \bibfield  {author} {\bibinfo {author} {\bibfnamefont {Z.}~\bibnamefont
  {Wang}}, \bibinfo {author} {\bibfnamefont {B.}~\bibnamefont {D'Anjou}},
  \bibinfo {author} {\bibfnamefont {P.}~\bibnamefont {Gigon}}, \bibinfo
  {author} {\bibfnamefont {A.}~\bibnamefont {Blais}},\ and\ \bibinfo {author}
  {\bibfnamefont {M.~S.}\ \bibnamefont {Blok}},\ }\href
  {https://doi.org/10.48550/arXiv.2505.00639} {\bibinfo {title} {Probing
  excited-state dynamics of transmon ionization}} (\bibinfo {year} {2025}),\
  \bibinfo {note} {arXiv:2505.00639 [quant-ph]}\BibitemShut {NoStop}%
\bibitem [{\citenamefont {Dai}\ \emph {et~al.}(2025)\citenamefont {Dai},
  \citenamefont {Hazra}, \citenamefont {Weiss}, \citenamefont {Kurilovich},
  \citenamefont {Connolly}, \citenamefont {Babla}, \citenamefont {Singh},
  \citenamefont {Joshi}, \citenamefont {Ding}, \citenamefont {Parakh},
  \citenamefont {Venkatraman}, \citenamefont {Xiao}, \citenamefont {Frunzio},\
  and\ \citenamefont {Devoret}}]{dai_spectroscopy_2025}%
  \BibitemOpen
  \bibfield  {author} {\bibinfo {author} {\bibfnamefont {W.}~\bibnamefont
  {Dai}}, \bibinfo {author} {\bibfnamefont {S.}~\bibnamefont {Hazra}}, \bibinfo
  {author} {\bibfnamefont {D.~K.}\ \bibnamefont {Weiss}}, \bibinfo {author}
  {\bibfnamefont {P.~D.}\ \bibnamefont {Kurilovich}}, \bibinfo {author}
  {\bibfnamefont {T.}~\bibnamefont {Connolly}}, \bibinfo {author}
  {\bibfnamefont {H.~K.}\ \bibnamefont {Babla}}, \bibinfo {author}
  {\bibfnamefont {S.}~\bibnamefont {Singh}}, \bibinfo {author} {\bibfnamefont
  {V.~R.}\ \bibnamefont {Joshi}}, \bibinfo {author} {\bibfnamefont {A.~Z.}\
  \bibnamefont {Ding}}, \bibinfo {author} {\bibfnamefont {P.~D.}\ \bibnamefont
  {Parakh}}, \bibinfo {author} {\bibfnamefont {J.}~\bibnamefont {Venkatraman}},
  \bibinfo {author} {\bibfnamefont {X.}~\bibnamefont {Xiao}}, \bibinfo {author}
  {\bibfnamefont {L.}~\bibnamefont {Frunzio}},\ and\ \bibinfo {author}
  {\bibfnamefont {M.~H.}\ \bibnamefont {Devoret}},\ }\href
  {https://doi.org/10.48550/arXiv.2506.24070} {\bibinfo {title} {Spectroscopy
  of drive-induced unwanted state transitions in superconducting circuits}}
  (\bibinfo {year} {2025}),\ \bibinfo {note} {arXiv:2506.24070
  [quant-ph]}\BibitemShut {NoStop}%
\bibitem [{\citenamefont {Shillito}\ \emph {et~al.}(2022)\citenamefont
  {Shillito}, \citenamefont {Petrescu}, \citenamefont {Cohen}, \citenamefont
  {Beall}, \citenamefont {Hauru}, \citenamefont {Ganahl}, \citenamefont
  {Lewis}, \citenamefont {Vidal},\ and\ \citenamefont {Blais}}]{Shillito2022}%
  \BibitemOpen
  \bibfield  {author} {\bibinfo {author} {\bibfnamefont {R.}~\bibnamefont
  {Shillito}}, \bibinfo {author} {\bibfnamefont {A.}~\bibnamefont {Petrescu}},
  \bibinfo {author} {\bibfnamefont {J.}~\bibnamefont {Cohen}}, \bibinfo
  {author} {\bibfnamefont {J.}~\bibnamefont {Beall}}, \bibinfo {author}
  {\bibfnamefont {M.}~\bibnamefont {Hauru}}, \bibinfo {author} {\bibfnamefont
  {M.}~\bibnamefont {Ganahl}}, \bibinfo {author} {\bibfnamefont {A.~G.}\
  \bibnamefont {Lewis}}, \bibinfo {author} {\bibfnamefont {G.}~\bibnamefont
  {Vidal}},\ and\ \bibinfo {author} {\bibfnamefont {A.}~\bibnamefont {Blais}},\
  }\bibfield  {title} {\bibinfo {title} {Dynamics of transmon ionization},\
  }\href {https://doi.org/10.1103/PhysRevApplied.18.034031} {\bibfield
  {journal} {\bibinfo  {journal} {Phys. Rev. Appl.}\ }\textbf {\bibinfo
  {volume} {18}},\ \bibinfo {pages} {034031} (\bibinfo {year}
  {2022})}\BibitemShut {NoStop}%
\bibitem [{\citenamefont {Nesterov}\ and\ \citenamefont
  {Pechenezhskiy}(2024)}]{Nesterov2024}%
  \BibitemOpen
  \bibfield  {author} {\bibinfo {author} {\bibfnamefont {K.~N.}\ \bibnamefont
  {Nesterov}}\ and\ \bibinfo {author} {\bibfnamefont {I.~V.}\ \bibnamefont
  {Pechenezhskiy}},\ }\bibfield  {title} {\bibinfo {title} {Measurement-induced
  state transitions in dispersive qubit-readout schemes},\ }\href
  {https://doi.org/10.1103/PhysRevApplied.22.064038} {\bibfield  {journal}
  {\bibinfo  {journal} {Phys. Rev. Appl.}\ }\textbf {\bibinfo {volume} {22}},\
  \bibinfo {pages} {064038} (\bibinfo {year} {2024})}\BibitemShut {NoStop}%
\bibitem [{\citenamefont {Singh}\ \emph {et~al.}(2024)\citenamefont {Singh},
  \citenamefont {Refael}, \citenamefont {Clerk},\ and\ \citenamefont
  {Rosenfeld}}]{Singh2024}%
  \BibitemOpen
  \bibfield  {author} {\bibinfo {author} {\bibfnamefont {S.}~\bibnamefont
  {Singh}}, \bibinfo {author} {\bibfnamefont {G.}~\bibnamefont {Refael}},
  \bibinfo {author} {\bibfnamefont {A.}~\bibnamefont {Clerk}},\ and\ \bibinfo
  {author} {\bibfnamefont {E.}~\bibnamefont {Rosenfeld}},\ }\bibfield  {title}
  {\bibinfo {title} {Impact of josephson junction array modes on fluxonium
  readout},\ }\href {https://arxiv.org/abs/2412.14788} {\bibfield  {journal}
  {\bibinfo  {journal} {arXiv preprint}\ } (\bibinfo {year} {2024})},\ \Eprint
  {https://arxiv.org/abs/2412.14788} {arXiv:2412.14788 [quant-ph]} \BibitemShut
  {NoStop}%
\bibitem [{\citenamefont {Stefanski}\ and\ \citenamefont
  {Andersen}(2024)}]{Stefanski2024}%
  \BibitemOpen
  \bibfield  {author} {\bibinfo {author} {\bibfnamefont {T.~V.}\ \bibnamefont
  {Stefanski}}\ and\ \bibinfo {author} {\bibfnamefont {C.~K.}\ \bibnamefont
  {Andersen}},\ }\bibfield  {title} {\bibinfo {title} {Flux-pulse-assisted
  readout of a fluxonium qubit},\ }\href
  {https://doi.org/10.1103/PhysRevApplied.22.014079} {\bibfield  {journal}
  {\bibinfo  {journal} {Phys. Rev. Appl.}\ }\textbf {\bibinfo {volume} {22}},\
  \bibinfo {pages} {014079} (\bibinfo {year} {2024})}\BibitemShut {NoStop}%
\bibitem [{\citenamefont {Clerk}\ and\ \citenamefont
  {Utami}(2007)}]{clerk_using_2007}%
  \BibitemOpen
  \bibfield  {author} {\bibinfo {author} {\bibfnamefont {A.~A.}\ \bibnamefont
  {Clerk}}\ and\ \bibinfo {author} {\bibfnamefont {D.~W.}\ \bibnamefont
  {Utami}},\ }\bibfield  {title} {\bibinfo {title} {Using a qubit to measure
  photon-number statistics of a driven thermal oscillator},\ }\href
  {https://doi.org/10.1103/PhysRevA.75.042302} {\bibfield  {journal} {\bibinfo
  {journal} {Physical Review A - Atomic, Molecular, and Optical Physics}\
  }\textbf {\bibinfo {volume} {75}},\ \bibinfo {pages} {1} (\bibinfo {year}
  {2007})}\BibitemShut {NoStop}%
\bibitem [{\citenamefont {Rigetti}\ \emph {et~al.}(2012)\citenamefont
  {Rigetti}, \citenamefont {Gambetta}, \citenamefont {Poletto}, \citenamefont
  {Plourde}, \citenamefont {Chow}, \citenamefont {Córcoles}, \citenamefont
  {Smolin}, \citenamefont {Merkel}, \citenamefont {Rozen}, \citenamefont
  {Keefe}, \citenamefont {Rothwell}, \citenamefont {Ketchen},\ and\
  \citenamefont {Steffen}}]{rigetti_superconducting_2012}%
  \BibitemOpen
  \bibfield  {author} {\bibinfo {author} {\bibfnamefont {C.}~\bibnamefont
  {Rigetti}}, \bibinfo {author} {\bibfnamefont {J.~M.}\ \bibnamefont
  {Gambetta}}, \bibinfo {author} {\bibfnamefont {S.}~\bibnamefont {Poletto}},
  \bibinfo {author} {\bibfnamefont {B.~L.}\ \bibnamefont {Plourde}}, \bibinfo
  {author} {\bibfnamefont {J.~M.}\ \bibnamefont {Chow}}, \bibinfo {author}
  {\bibfnamefont {A.~D.}\ \bibnamefont {Córcoles}}, \bibinfo {author}
  {\bibfnamefont {J.~A.}\ \bibnamefont {Smolin}}, \bibinfo {author}
  {\bibfnamefont {S.~T.}\ \bibnamefont {Merkel}}, \bibinfo {author}
  {\bibfnamefont {J.~R.}\ \bibnamefont {Rozen}}, \bibinfo {author}
  {\bibfnamefont {G.~A.}\ \bibnamefont {Keefe}}, \bibinfo {author}
  {\bibfnamefont {M.~B.}\ \bibnamefont {Rothwell}}, \bibinfo {author}
  {\bibfnamefont {M.~B.}\ \bibnamefont {Ketchen}},\ and\ \bibinfo {author}
  {\bibfnamefont {M.}~\bibnamefont {Steffen}},\ }\bibfield  {title} {\bibinfo
  {title} {Superconducting qubit in a waveguide cavity with a coherence time
  approaching 0.1 ms},\ }\href {https://doi.org/10.1103/PhysRevB.86.100506}
  {\bibfield  {journal} {\bibinfo  {journal} {Physical Review B - Condensed
  Matter and Materials Physics}\ }\textbf {\bibinfo {volume} {86}},\ \bibinfo
  {pages} {1} (\bibinfo {year} {2012})},\ \bibinfo {note} {arXiv:
  1202.5533}\BibitemShut {NoStop}%
\bibitem [{\citenamefont {Zhang}\ \emph {et~al.}(2017)\citenamefont {Zhang},
  \citenamefont {Liu}, \citenamefont {Raftery},\ and\ \citenamefont
  {Houck}}]{zhang_suppression_2017}%
  \BibitemOpen
  \bibfield  {author} {\bibinfo {author} {\bibfnamefont {G.}~\bibnamefont
  {Zhang}}, \bibinfo {author} {\bibfnamefont {Y.}~\bibnamefont {Liu}}, \bibinfo
  {author} {\bibfnamefont {J.~J.}\ \bibnamefont {Raftery}},\ and\ \bibinfo
  {author} {\bibfnamefont {A.~A.}\ \bibnamefont {Houck}},\ }\bibfield  {title}
  {\bibinfo {title} {Suppression of photon shot noise dephasing in a tunable
  coupling superconducting qubit},\ }\bibfield  {journal} {\bibinfo  {journal}
  {npj Quantum Information}\ }\textbf {\bibinfo {volume} {3}},\ \href
  {https://doi.org/10.1038/s41534-016-0002-2} {10.1038/s41534-016-0002-2}
  (\bibinfo {year} {2017}),\ \bibinfo {note} {arXiv: 1603.01224}\BibitemShut
  {NoStop}%
\bibitem [{\citenamefont {Wang}\ \emph {et~al.}(2019)\citenamefont {Wang},
  \citenamefont {Shankar}, \citenamefont {Minev}, \citenamefont
  {Campagne-Ibarcq}, \citenamefont {Narla},\ and\ \citenamefont
  {Devoret}}]{wang_cavity_2019}%
  \BibitemOpen
  \bibfield  {author} {\bibinfo {author} {\bibfnamefont {Z.}~\bibnamefont
  {Wang}}, \bibinfo {author} {\bibfnamefont {S.}~\bibnamefont {Shankar}},
  \bibinfo {author} {\bibfnamefont {Z.~K.}\ \bibnamefont {Minev}}, \bibinfo
  {author} {\bibfnamefont {P.}~\bibnamefont {Campagne-Ibarcq}}, \bibinfo
  {author} {\bibfnamefont {A.}~\bibnamefont {Narla}},\ and\ \bibinfo {author}
  {\bibfnamefont {M.~H.}\ \bibnamefont {Devoret}},\ }\bibfield  {title}
  {\bibinfo {title} {Cavity {Attenuators} for {Superconducting} {Qubits}},\
  }\href {https://doi.org/10.1103/PhysRevApplied.11.014031} {\bibfield
  {journal} {\bibinfo  {journal} {Physical Review Applied}\ }\textbf {\bibinfo
  {volume} {11}},\ \bibinfo {pages} {1} (\bibinfo {year} {2019})},\ \bibinfo
  {note} {arXiv: 1807.04849 Publisher: American Physical Society}\BibitemShut
  {NoStop}%
\bibitem [{\citenamefont {Barends}\ \emph {et~al.}(2013)\citenamefont
  {Barends}, \citenamefont {Kelly}, \citenamefont {Megrant}, \citenamefont
  {Sank}, \citenamefont {Jeffrey}, \citenamefont {Chen}, \citenamefont {Yin},
  \citenamefont {Chiaro}, \citenamefont {Mutus}, \citenamefont {Neill},
  \citenamefont {O’Malley}, \citenamefont {Roushan}, \citenamefont {Wenner},
  \citenamefont {White}, \citenamefont {Cleland},\ and\ \citenamefont
  {Martinis}}]{barends_coherent_2013}%
  \BibitemOpen
  \bibfield  {author} {\bibinfo {author} {\bibfnamefont {R.}~\bibnamefont
  {Barends}}, \bibinfo {author} {\bibfnamefont {J.}~\bibnamefont {Kelly}},
  \bibinfo {author} {\bibfnamefont {A.}~\bibnamefont {Megrant}}, \bibinfo
  {author} {\bibfnamefont {D.}~\bibnamefont {Sank}}, \bibinfo {author}
  {\bibfnamefont {E.}~\bibnamefont {Jeffrey}}, \bibinfo {author} {\bibfnamefont
  {Y.}~\bibnamefont {Chen}}, \bibinfo {author} {\bibfnamefont {Y.}~\bibnamefont
  {Yin}}, \bibinfo {author} {\bibfnamefont {B.}~\bibnamefont {Chiaro}},
  \bibinfo {author} {\bibfnamefont {J.}~\bibnamefont {Mutus}}, \bibinfo
  {author} {\bibfnamefont {C.}~\bibnamefont {Neill}}, \bibinfo {author}
  {\bibfnamefont {P.}~\bibnamefont {O’Malley}}, \bibinfo {author}
  {\bibfnamefont {P.}~\bibnamefont {Roushan}}, \bibinfo {author} {\bibfnamefont
  {J.}~\bibnamefont {Wenner}}, \bibinfo {author} {\bibfnamefont {T.~C.}\
  \bibnamefont {White}}, \bibinfo {author} {\bibfnamefont {A.~N.}\ \bibnamefont
  {Cleland}},\ and\ \bibinfo {author} {\bibfnamefont {J.~M.}\ \bibnamefont
  {Martinis}},\ }\bibfield  {title} {\bibinfo {title} {Coherent {Josephson}
  {Qubit} {Suitable} for {Scalable} {Quantum} {Integrated} {Circuits}},\ }\href
  {https://doi.org/10.1103/PhysRevLett.111.080502} {\bibfield  {journal}
  {\bibinfo  {journal} {Physical Review Letters}\ }\textbf {\bibinfo {volume}
  {111}},\ \bibinfo {pages} {080502} (\bibinfo {year} {2013})}\BibitemShut
  {NoStop}%
\bibitem [{\citenamefont {Thorbeck}\ \emph {et~al.}(2024)\citenamefont
  {Thorbeck}, \citenamefont {Xiao}, \citenamefont {Kamal},\ and\ \citenamefont
  {Govia}}]{thorbeck_readout-induced_2024}%
  \BibitemOpen
  \bibfield  {author} {\bibinfo {author} {\bibfnamefont {T.}~\bibnamefont
  {Thorbeck}}, \bibinfo {author} {\bibfnamefont {Z.}~\bibnamefont {Xiao}},
  \bibinfo {author} {\bibfnamefont {A.}~\bibnamefont {Kamal}},\ and\ \bibinfo
  {author} {\bibfnamefont {L.~C.}\ \bibnamefont {Govia}},\ }\bibfield  {title}
  {\bibinfo {title} {Readout-{Induced} {Suppression} and {Enhancement} of
  {Superconducting} {Qubit} {Lifetimes}},\ }\href
  {https://doi.org/10.1103/PhysRevLett.132.090602} {\bibfield  {journal}
  {\bibinfo  {journal} {Physical Review Letters}\ }\textbf {\bibinfo {volume}
  {132}},\ \bibinfo {pages} {090602} (\bibinfo {year} {2024})}\BibitemShut
  {NoStop}%
\bibitem [{\citenamefont {Manucharyan}\ \emph {et~al.}(2009)\citenamefont
  {Manucharyan}, \citenamefont {Koch}, \citenamefont {Glazman},\ and\
  \citenamefont {Devoret}}]{manucharyan_fluxonium_2009}%
  \BibitemOpen
  \bibfield  {author} {\bibinfo {author} {\bibfnamefont {V.~E.}\ \bibnamefont
  {Manucharyan}}, \bibinfo {author} {\bibfnamefont {J.}~\bibnamefont {Koch}},
  \bibinfo {author} {\bibfnamefont {L.~I.}\ \bibnamefont {Glazman}},\ and\
  \bibinfo {author} {\bibfnamefont {M.~H.}\ \bibnamefont {Devoret}},\
  }\bibfield  {title} {\bibinfo {title} {Fluxonium: {Single} cooper-pair
  circuit free of charge offsets},\ }\href
  {https://doi.org/10.1126/science.1175552} {\bibfield  {journal} {\bibinfo
  {journal} {Science}\ }\textbf {\bibinfo {volume} {326}},\ \bibinfo {pages}
  {113} (\bibinfo {year} {2009})},\ \bibinfo {note} {arXiv:
  0906.0831}\BibitemShut {NoStop}%
\bibitem [{\citenamefont {Ding}\ \emph {et~al.}(2023)\citenamefont {Ding},
  \citenamefont {Hays}, \citenamefont {Sung}, \citenamefont {Kannan},
  \citenamefont {An}, \citenamefont {Di~Paolo}, \citenamefont {Karamlou},
  \citenamefont {Hazard}, \citenamefont {Azar}, \citenamefont {Kim},
  \citenamefont {Niedzielski}, \citenamefont {Melville}, \citenamefont
  {Schwartz}, \citenamefont {Yoder}, \citenamefont {Orlando}, \citenamefont
  {Gustavsson}, \citenamefont {Grover}, \citenamefont {Serniak},\ and\
  \citenamefont {Oliver}}]{ding_high-fidelity_2023}%
  \BibitemOpen
  \bibfield  {author} {\bibinfo {author} {\bibfnamefont {L.}~\bibnamefont
  {Ding}}, \bibinfo {author} {\bibfnamefont {M.}~\bibnamefont {Hays}}, \bibinfo
  {author} {\bibfnamefont {Y.}~\bibnamefont {Sung}}, \bibinfo {author}
  {\bibfnamefont {B.}~\bibnamefont {Kannan}}, \bibinfo {author} {\bibfnamefont
  {J.}~\bibnamefont {An}}, \bibinfo {author} {\bibfnamefont {A.}~\bibnamefont
  {Di~Paolo}}, \bibinfo {author} {\bibfnamefont {A.~H.}\ \bibnamefont
  {Karamlou}}, \bibinfo {author} {\bibfnamefont {T.~M.}\ \bibnamefont
  {Hazard}}, \bibinfo {author} {\bibfnamefont {K.}~\bibnamefont {Azar}},
  \bibinfo {author} {\bibfnamefont {D.~K.}\ \bibnamefont {Kim}}, \bibinfo
  {author} {\bibfnamefont {B.~M.}\ \bibnamefont {Niedzielski}}, \bibinfo
  {author} {\bibfnamefont {A.}~\bibnamefont {Melville}}, \bibinfo {author}
  {\bibfnamefont {M.~E.}\ \bibnamefont {Schwartz}}, \bibinfo {author}
  {\bibfnamefont {J.~L.}\ \bibnamefont {Yoder}}, \bibinfo {author}
  {\bibfnamefont {T.~P.}\ \bibnamefont {Orlando}}, \bibinfo {author}
  {\bibfnamefont {S.}~\bibnamefont {Gustavsson}}, \bibinfo {author}
  {\bibfnamefont {J.~A.}\ \bibnamefont {Grover}}, \bibinfo {author}
  {\bibfnamefont {K.}~\bibnamefont {Serniak}},\ and\ \bibinfo {author}
  {\bibfnamefont {W.~D.}\ \bibnamefont {Oliver}},\ }\bibfield  {title}
  {\bibinfo {title} {High-{Fidelity}, {Frequency}-{Flexible} {Two}-{Qubit}
  {Fluxonium} {Gates} with a {Transmon} {Coupler}},\ }\href
  {https://doi.org/10.1103/PhysRevX.13.031035} {\bibfield  {journal} {\bibinfo
  {journal} {Physical Review X}\ }\textbf {\bibinfo {volume} {13}},\ \bibinfo
  {pages} {031035} (\bibinfo {year} {2023})}\BibitemShut {NoStop}%
\bibitem [{\citenamefont {Xiong}\ \emph {et~al.}(2025)\citenamefont {Xiong},
  \citenamefont {Wang}, \citenamefont {Song}, \citenamefont {Yang},
  \citenamefont {Bao}, \citenamefont {Li}, \citenamefont {Mi}, \citenamefont
  {Zhang}, \citenamefont {Yu}, \citenamefont {Song},\ and\ \citenamefont
  {Duan}}]{xiong_scalable_2025}%
  \BibitemOpen
  \bibfield  {author} {\bibinfo {author} {\bibfnamefont {H.}~\bibnamefont
  {Xiong}}, \bibinfo {author} {\bibfnamefont {J.}~\bibnamefont {Wang}},
  \bibinfo {author} {\bibfnamefont {J.}~\bibnamefont {Song}}, \bibinfo {author}
  {\bibfnamefont {J.}~\bibnamefont {Yang}}, \bibinfo {author} {\bibfnamefont
  {Z.}~\bibnamefont {Bao}}, \bibinfo {author} {\bibfnamefont {Y.}~\bibnamefont
  {Li}}, \bibinfo {author} {\bibfnamefont {Z.-Y.}\ \bibnamefont {Mi}}, \bibinfo
  {author} {\bibfnamefont {H.}~\bibnamefont {Zhang}}, \bibinfo {author}
  {\bibfnamefont {H.-F.}\ \bibnamefont {Yu}}, \bibinfo {author} {\bibfnamefont
  {Y.}~\bibnamefont {Song}},\ and\ \bibinfo {author} {\bibfnamefont
  {L.}~\bibnamefont {Duan}},\ }\href
  {https://doi.org/10.48550/arXiv.2502.18902} {\bibinfo {title} {Scalable
  {Low}-overhead {Superconducting} {Non}-local {Coupler} with {Exponentially}
  {Enhanced} {Connectivity}}} (\bibinfo {year} {2025}),\ \bibinfo {note}
  {arXiv:2502.18902 [quant-ph]}\BibitemShut {NoStop}%
\bibitem [{\citenamefont {Wang}\ \emph {et~al.}(2024)\citenamefont {Wang},
  \citenamefont {Wu}, \citenamefont {Wang}, \citenamefont {Ma}, \citenamefont
  {Zhang}, \citenamefont {Chen}, \citenamefont {Deng}, \citenamefont {Gao},
  \citenamefont {Hu}, \citenamefont {Ma}, \citenamefont {Song}, \citenamefont
  {Xia}, \citenamefont {Ying}, \citenamefont {Zhan}, \citenamefont {Zhao},\
  and\ \citenamefont {Deng}}]{wang_efficient_2024}%
  \BibitemOpen
  \bibfield  {author} {\bibinfo {author} {\bibfnamefont {T.}~\bibnamefont
  {Wang}}, \bibinfo {author} {\bibfnamefont {F.}~\bibnamefont {Wu}}, \bibinfo
  {author} {\bibfnamefont {F.}~\bibnamefont {Wang}}, \bibinfo {author}
  {\bibfnamefont {X.}~\bibnamefont {Ma}}, \bibinfo {author} {\bibfnamefont
  {G.}~\bibnamefont {Zhang}}, \bibinfo {author} {\bibfnamefont
  {J.}~\bibnamefont {Chen}}, \bibinfo {author} {\bibfnamefont {H.}~\bibnamefont
  {Deng}}, \bibinfo {author} {\bibfnamefont {R.}~\bibnamefont {Gao}}, \bibinfo
  {author} {\bibfnamefont {R.}~\bibnamefont {Hu}}, \bibinfo {author}
  {\bibfnamefont {L.}~\bibnamefont {Ma}}, \bibinfo {author} {\bibfnamefont
  {Z.}~\bibnamefont {Song}}, \bibinfo {author} {\bibfnamefont {T.}~\bibnamefont
  {Xia}}, \bibinfo {author} {\bibfnamefont {M.}~\bibnamefont {Ying}}, \bibinfo
  {author} {\bibfnamefont {H.}~\bibnamefont {Zhan}}, \bibinfo {author}
  {\bibfnamefont {H.-H.}\ \bibnamefont {Zhao}},\ and\ \bibinfo {author}
  {\bibfnamefont {C.}~\bibnamefont {Deng}},\ }\href
  {http://arxiv.org/abs/2402.06267} {\bibinfo {title} {Efficient initialization
  of fluxonium qubits based on auxiliary energy levels}} (\bibinfo {year}
  {2024}),\ \bibinfo {note} {arXiv:2402.06267 [quant-ph]}\BibitemShut {NoStop}%
\bibitem [{\citenamefont {Kuzmin}\ \emph {et~al.}(2019)\citenamefont {Kuzmin},
  \citenamefont {Mehta}, \citenamefont {Grabon}, \citenamefont {Mencia},\ and\
  \citenamefont {Manucharyan}}]{kuzmin_superstrong_2019}%
  \BibitemOpen
  \bibfield  {author} {\bibinfo {author} {\bibfnamefont {R.}~\bibnamefont
  {Kuzmin}}, \bibinfo {author} {\bibfnamefont {N.}~\bibnamefont {Mehta}},
  \bibinfo {author} {\bibfnamefont {N.}~\bibnamefont {Grabon}}, \bibinfo
  {author} {\bibfnamefont {R.}~\bibnamefont {Mencia}},\ and\ \bibinfo {author}
  {\bibfnamefont {V.~E.}\ \bibnamefont {Manucharyan}},\ }\bibfield  {title}
  {\bibinfo {title} {Superstrong coupling in circuit quantum electrodynamics},\
  }\href {https://doi.org/10.1038/s41534-019-0134-2} {\bibfield  {journal}
  {\bibinfo  {journal} {npj Quantum Information}\ }\textbf {\bibinfo {volume}
  {5}},\ \bibinfo {pages} {20} (\bibinfo {year} {2019})},\ \bibinfo {note}
  {arXiv: 1809.10739 Publisher: Springer US}\BibitemShut {NoStop}%
\bibitem [{\citenamefont {Mehta}(2022)}]{mehta_quantum_2022}%
  \BibitemOpen
  \bibfield  {author} {\bibinfo {author} {\bibfnamefont {N.}~\bibnamefont
  {Mehta}},\ }\bibfield  {title} {\bibinfo {title} {Quantum {Impurity} {Regime}
  of {Circuit} {Quantum} {Electrodynamics}},\ }\href@noop {} {\  (\bibinfo
  {year} {2022})}\BibitemShut {NoStop}%
\bibitem [{\citenamefont {Mencia}(2023)}]{mencia_ultra-high_2023}%
  \BibitemOpen
  \bibfield  {author} {\bibinfo {author} {\bibfnamefont {R.~A.}\ \bibnamefont
  {Mencia}},\ }\bibfield  {title} {\bibinfo {title} {Ultra-high impedance
  superconducting circuits},\ }\href@noop {} {\  (\bibinfo {year}
  {2023})}\BibitemShut {NoStop}%
\bibitem [{\citenamefont {Zhuang}\ \emph {et~al.}(2025)\citenamefont {Zhuang},
  \citenamefont {Rosenstock}, \citenamefont {Liu}, \citenamefont {Somoroff},
  \citenamefont {Manucharyan},\ and\ \citenamefont
  {Wang}}]{zhuang_non-markovian_2025}%
  \BibitemOpen
  \bibfield  {author} {\bibinfo {author} {\bibfnamefont {Z.-T.}\ \bibnamefont
  {Zhuang}}, \bibinfo {author} {\bibfnamefont {D.}~\bibnamefont {Rosenstock}},
  \bibinfo {author} {\bibfnamefont {B.-J.}\ \bibnamefont {Liu}}, \bibinfo
  {author} {\bibfnamefont {A.}~\bibnamefont {Somoroff}}, \bibinfo {author}
  {\bibfnamefont {V.~E.}\ \bibnamefont {Manucharyan}},\ and\ \bibinfo {author}
  {\bibfnamefont {C.}~\bibnamefont {Wang}},\ }\href
  {https://doi.org/10.48550/arXiv.2503.16381} {\bibinfo {title}
  {Non-{Markovian} {Relaxation} {Spectroscopy} of {Fluxonium} {Qubits}}}
  (\bibinfo {year} {2025}),\ \bibinfo {note} {arXiv:2503.16381
  [quant-ph]}\BibitemShut {NoStop}%
\bibitem [{\citenamefont {McClure}\ \emph {et~al.}(2016)\citenamefont
  {McClure}, \citenamefont {Paik}, \citenamefont {Bishop}, \citenamefont
  {Steffen}, \citenamefont {Chow},\ and\ \citenamefont
  {Gambetta}}]{mcclure_rapid_2016}%
  \BibitemOpen
  \bibfield  {author} {\bibinfo {author} {\bibfnamefont {D.}~\bibnamefont
  {McClure}}, \bibinfo {author} {\bibfnamefont {H.}~\bibnamefont {Paik}},
  \bibinfo {author} {\bibfnamefont {L.}~\bibnamefont {Bishop}}, \bibinfo
  {author} {\bibfnamefont {M.}~\bibnamefont {Steffen}}, \bibinfo {author}
  {\bibfnamefont {J.~M.}\ \bibnamefont {Chow}},\ and\ \bibinfo {author}
  {\bibfnamefont {J.~M.}\ \bibnamefont {Gambetta}},\ }\bibfield  {title}
  {\bibinfo {title} {Rapid {Driven} {Reset} of a {Qubit} {Readout}
  {Resonator}},\ }\href {https://doi.org/10.1103/PhysRevApplied.5.011001}
  {\bibfield  {journal} {\bibinfo  {journal} {Physical Review Applied}\
  }\textbf {\bibinfo {volume} {5}},\ \bibinfo {pages} {011001} (\bibinfo {year}
  {2016})}\BibitemShut {NoStop}%
\bibitem [{\citenamefont {Bultink}\ \emph {et~al.}(2018)\citenamefont
  {Bultink}, \citenamefont {Tarasinski}, \citenamefont {Haandbaek},
  \citenamefont {Poletto}, \citenamefont {Haider}, \citenamefont {Michalak},
  \citenamefont {Bruno},\ and\ \citenamefont {DiCarlo}}]{bultink_general_2018}%
  \BibitemOpen
  \bibfield  {author} {\bibinfo {author} {\bibfnamefont {C.~C.}\ \bibnamefont
  {Bultink}}, \bibinfo {author} {\bibfnamefont {B.}~\bibnamefont {Tarasinski}},
  \bibinfo {author} {\bibfnamefont {N.}~\bibnamefont {Haandbaek}}, \bibinfo
  {author} {\bibfnamefont {S.}~\bibnamefont {Poletto}}, \bibinfo {author}
  {\bibfnamefont {N.}~\bibnamefont {Haider}}, \bibinfo {author} {\bibfnamefont
  {D.~J.}\ \bibnamefont {Michalak}}, \bibinfo {author} {\bibfnamefont
  {A.}~\bibnamefont {Bruno}},\ and\ \bibinfo {author} {\bibfnamefont
  {L.}~\bibnamefont {DiCarlo}},\ }\bibfield  {title} {\bibinfo {title} {General
  method for extracting the quantum efficiency of dispersive qubit readout in
  circuit {QED}},\ }\href {https://doi.org/10.1063/1.5015954} {\bibfield
  {journal} {\bibinfo  {journal} {Applied Physics Letters}\ }\textbf {\bibinfo
  {volume} {112}},\ \bibinfo {pages} {092601} (\bibinfo {year} {2018})},\
  \bibinfo {note} {arXiv:1711.05336 [quant-ph]}\BibitemShut {NoStop}%
\bibitem [{\citenamefont {Jerger}\ \emph {et~al.}(2024)\citenamefont {Jerger},
  \citenamefont {Motzoi}, \citenamefont {Gao}, \citenamefont {Dickel},
  \citenamefont {Buchmann}, \citenamefont {Bengtsson}, \citenamefont
  {Tancredi}, \citenamefont {Warren}, \citenamefont {Bylander}, \citenamefont
  {DiVincenzo}, \citenamefont {Barends},\ and\ \citenamefont
  {Bushev}}]{jerger_dispersive_2024}%
  \BibitemOpen
  \bibfield  {author} {\bibinfo {author} {\bibfnamefont {M.}~\bibnamefont
  {Jerger}}, \bibinfo {author} {\bibfnamefont {F.}~\bibnamefont {Motzoi}},
  \bibinfo {author} {\bibfnamefont {Y.}~\bibnamefont {Gao}}, \bibinfo {author}
  {\bibfnamefont {C.}~\bibnamefont {Dickel}}, \bibinfo {author} {\bibfnamefont
  {L.}~\bibnamefont {Buchmann}}, \bibinfo {author} {\bibfnamefont
  {A.}~\bibnamefont {Bengtsson}}, \bibinfo {author} {\bibfnamefont
  {G.}~\bibnamefont {Tancredi}}, \bibinfo {author} {\bibfnamefont {C.~W.}\
  \bibnamefont {Warren}}, \bibinfo {author} {\bibfnamefont {J.}~\bibnamefont
  {Bylander}}, \bibinfo {author} {\bibfnamefont {D.}~\bibnamefont
  {DiVincenzo}}, \bibinfo {author} {\bibfnamefont {R.}~\bibnamefont
  {Barends}},\ and\ \bibinfo {author} {\bibfnamefont {P.~A.}\ \bibnamefont
  {Bushev}},\ }\href {https://doi.org/10.48550/arXiv.2406.04891} {\bibinfo
  {title} {Dispersive {Qubit} {Readout} with {Intrinsic} {Resonator} {Reset}}}
  (\bibinfo {year} {2024}),\ \bibinfo {note} {arXiv:2406.04891
  [quant-ph]}\BibitemShut {NoStop}%
\bibitem [{\citenamefont {Klimov}\ \emph {et~al.}(2018)\citenamefont {Klimov},
  \citenamefont {Kelly}, \citenamefont {Chen}, \citenamefont {Neeley},
  \citenamefont {Megrant}, \citenamefont {Burkett}, \citenamefont {Barends},
  \citenamefont {Arya}, \citenamefont {Chiaro}, \citenamefont {Chen},
  \citenamefont {Dunsworth}, \citenamefont {Fowler}, \citenamefont {Foxen},
  \citenamefont {Gidney}, \citenamefont {Giustina}, \citenamefont {Graff},
  \citenamefont {Huang}, \citenamefont {Jeffrey}, \citenamefont {Lucero},
  \citenamefont {Mutus}, \citenamefont {Naaman}, \citenamefont {Neill},
  \citenamefont {Quintana}, \citenamefont {Roushan}, \citenamefont {Sank},
  \citenamefont {Vainsencher}, \citenamefont {Wenner}, \citenamefont {White},
  \citenamefont {Boixo}, \citenamefont {Babbush}, \citenamefont {Smelyanskiy},
  \citenamefont {Neven},\ and\ \citenamefont
  {Martinis}}]{klimov_fluctuations_2018}%
  \BibitemOpen
  \bibfield  {author} {\bibinfo {author} {\bibfnamefont {P.~V.}\ \bibnamefont
  {Klimov}}, \bibinfo {author} {\bibfnamefont {J.}~\bibnamefont {Kelly}},
  \bibinfo {author} {\bibfnamefont {Z.}~\bibnamefont {Chen}}, \bibinfo {author}
  {\bibfnamefont {M.}~\bibnamefont {Neeley}}, \bibinfo {author} {\bibfnamefont
  {A.}~\bibnamefont {Megrant}}, \bibinfo {author} {\bibfnamefont
  {B.}~\bibnamefont {Burkett}}, \bibinfo {author} {\bibfnamefont
  {R.}~\bibnamefont {Barends}}, \bibinfo {author} {\bibfnamefont
  {K.}~\bibnamefont {Arya}}, \bibinfo {author} {\bibfnamefont {B.}~\bibnamefont
  {Chiaro}}, \bibinfo {author} {\bibfnamefont {Y.}~\bibnamefont {Chen}},
  \bibinfo {author} {\bibfnamefont {A.}~\bibnamefont {Dunsworth}}, \bibinfo
  {author} {\bibfnamefont {A.}~\bibnamefont {Fowler}}, \bibinfo {author}
  {\bibfnamefont {B.}~\bibnamefont {Foxen}}, \bibinfo {author} {\bibfnamefont
  {C.}~\bibnamefont {Gidney}}, \bibinfo {author} {\bibfnamefont
  {M.}~\bibnamefont {Giustina}}, \bibinfo {author} {\bibfnamefont
  {R.}~\bibnamefont {Graff}}, \bibinfo {author} {\bibfnamefont
  {T.}~\bibnamefont {Huang}}, \bibinfo {author} {\bibfnamefont
  {E.}~\bibnamefont {Jeffrey}}, \bibinfo {author} {\bibfnamefont
  {E.}~\bibnamefont {Lucero}}, \bibinfo {author} {\bibfnamefont {J.~Y.}\
  \bibnamefont {Mutus}}, \bibinfo {author} {\bibfnamefont {O.}~\bibnamefont
  {Naaman}}, \bibinfo {author} {\bibfnamefont {C.}~\bibnamefont {Neill}},
  \bibinfo {author} {\bibfnamefont {C.}~\bibnamefont {Quintana}}, \bibinfo
  {author} {\bibfnamefont {P.}~\bibnamefont {Roushan}}, \bibinfo {author}
  {\bibfnamefont {D.}~\bibnamefont {Sank}}, \bibinfo {author} {\bibfnamefont
  {A.}~\bibnamefont {Vainsencher}}, \bibinfo {author} {\bibfnamefont
  {J.}~\bibnamefont {Wenner}}, \bibinfo {author} {\bibfnamefont {T.~C.}\
  \bibnamefont {White}}, \bibinfo {author} {\bibfnamefont {S.}~\bibnamefont
  {Boixo}}, \bibinfo {author} {\bibfnamefont {R.}~\bibnamefont {Babbush}},
  \bibinfo {author} {\bibfnamefont {V.~N.}\ \bibnamefont {Smelyanskiy}},
  \bibinfo {author} {\bibfnamefont {H.}~\bibnamefont {Neven}},\ and\ \bibinfo
  {author} {\bibfnamefont {J.~M.}\ \bibnamefont {Martinis}},\ }\bibfield
  {title} {\bibinfo {title} {Fluctuations of {Energy}-{Relaxation} {Times} in
  {Superconducting} {Qubits}},\ }\href
  {https://doi.org/10.1103/PhysRevLett.121.090502} {\bibfield  {journal}
  {\bibinfo  {journal} {Physical Review Letters}\ }\textbf {\bibinfo {volume}
  {121}},\ \bibinfo {pages} {90502} (\bibinfo {year} {2018})},\ \bibinfo {note}
  {publisher: American Physical Society}\BibitemShut {NoStop}%
\bibitem [{\citenamefont {Thorbeck}\ \emph {et~al.}(2023)\citenamefont
  {Thorbeck}, \citenamefont {Eddins}, \citenamefont {Lauer}, \citenamefont
  {McClure},\ and\ \citenamefont {Carroll}}]{thorbeck_two-level-system_2023}%
  \BibitemOpen
  \bibfield  {author} {\bibinfo {author} {\bibfnamefont {T.}~\bibnamefont
  {Thorbeck}}, \bibinfo {author} {\bibfnamefont {A.}~\bibnamefont {Eddins}},
  \bibinfo {author} {\bibfnamefont {I.}~\bibnamefont {Lauer}}, \bibinfo
  {author} {\bibfnamefont {D.~T.}\ \bibnamefont {McClure}},\ and\ \bibinfo
  {author} {\bibfnamefont {M.}~\bibnamefont {Carroll}},\ }\bibfield  {title}
  {\bibinfo {title} {Two-{Level}-{System} {Dynamics} in a {Superconducting}
  {Qubit} {Due} to {Background} {Ionizing} {Radiation}},\ }\href
  {https://doi.org/10.1103/PRXQuantum.4.020356} {\bibfield  {journal} {\bibinfo
   {journal} {PRX Quantum}\ }\textbf {\bibinfo {volume} {4}},\ \bibinfo {pages}
  {020356} (\bibinfo {year} {2023})}\BibitemShut {NoStop}%
\bibitem [{\citenamefont {Kurilovich}\ \emph {et~al.}(2025)\citenamefont
  {Kurilovich}, \citenamefont {Connolly}, \citenamefont {Bøttcher},
  \citenamefont {Weiss}, \citenamefont {Hazra}, \citenamefont {Joshi},
  \citenamefont {Ding}, \citenamefont {Nho}, \citenamefont {Diamond},
  \citenamefont {Kurilovich}, \citenamefont {Dai}, \citenamefont {Fatemi},
  \citenamefont {Frunzio}, \citenamefont {Glazman},\ and\ \citenamefont
  {Devoret}}]{kurilovich_high-frequency_2025}%
  \BibitemOpen
  \bibfield  {author} {\bibinfo {author} {\bibfnamefont {P.~D.}\ \bibnamefont
  {Kurilovich}}, \bibinfo {author} {\bibfnamefont {T.}~\bibnamefont
  {Connolly}}, \bibinfo {author} {\bibfnamefont {C.~G.~L.}\ \bibnamefont
  {Bøttcher}}, \bibinfo {author} {\bibfnamefont {D.~K.}\ \bibnamefont
  {Weiss}}, \bibinfo {author} {\bibfnamefont {S.}~\bibnamefont {Hazra}},
  \bibinfo {author} {\bibfnamefont {V.~R.}\ \bibnamefont {Joshi}}, \bibinfo
  {author} {\bibfnamefont {A.~Z.}\ \bibnamefont {Ding}}, \bibinfo {author}
  {\bibfnamefont {H.}~\bibnamefont {Nho}}, \bibinfo {author} {\bibfnamefont
  {S.}~\bibnamefont {Diamond}}, \bibinfo {author} {\bibfnamefont {V.~D.}\
  \bibnamefont {Kurilovich}}, \bibinfo {author} {\bibfnamefont
  {W.}~\bibnamefont {Dai}}, \bibinfo {author} {\bibfnamefont {V.}~\bibnamefont
  {Fatemi}}, \bibinfo {author} {\bibfnamefont {L.}~\bibnamefont {Frunzio}},
  \bibinfo {author} {\bibfnamefont {L.~I.}\ \bibnamefont {Glazman}},\ and\
  \bibinfo {author} {\bibfnamefont {M.~H.}\ \bibnamefont {Devoret}},\ }\href
  {https://doi.org/10.48550/arXiv.2501.09161} {\bibinfo {title} {High-frequency
  readout free from transmon multi-excitation resonances}} (\bibinfo {year}
  {2025}),\ \bibinfo {note} {arXiv:2501.09161 [quant-ph]}\BibitemShut {NoStop}%
\bibitem [{\citenamefont
  {Castellanos-Beltran}(2010)}]{castellanos-beltran_development_2010}%
  \BibitemOpen
  \bibfield  {author} {\bibinfo {author} {\bibfnamefont {M.~a.}\ \bibnamefont
  {Castellanos-Beltran}},\ }\bibfield  {title} {\bibinfo {title} {Development
  of a {Josephson} {Parametric} {Amplifier} for the {Preparation} and
  {Detection} of {Nonclassical} {States} of {Microwave} {Fields}},\ }\href@noop
  {} {\  (\bibinfo {year} {2010})}\BibitemShut {NoStop}%
\bibitem [{\citenamefont {Weber}(2014)}]{weber_quantum_2014}%
  \BibitemOpen
  \bibfield  {author} {\bibinfo {author} {\bibfnamefont {S.~J.}\ \bibnamefont
  {Weber}},\ }\bibfield  {title} {\bibinfo {title} {Quantum {Trajectories} of a
  {Superconducting} {Qubit}},\ }\href@noop {} {\  (\bibinfo {year}
  {2014})}\BibitemShut {NoStop}%
\bibitem [{\citenamefont {Touzard}\ \emph {et~al.}(2019)\citenamefont
  {Touzard}, \citenamefont {Kou}, \citenamefont {Frattini}, \citenamefont
  {Sivak}, \citenamefont {Puri}, \citenamefont {Grimm}, \citenamefont
  {Frunzio}, \citenamefont {Shankar},\ and\ \citenamefont
  {Devoret}}]{touzard_gated_2019}%
  \BibitemOpen
  \bibfield  {author} {\bibinfo {author} {\bibfnamefont {S.}~\bibnamefont
  {Touzard}}, \bibinfo {author} {\bibfnamefont {A.}~\bibnamefont {Kou}},
  \bibinfo {author} {\bibfnamefont {N.~E.}\ \bibnamefont {Frattini}}, \bibinfo
  {author} {\bibfnamefont {V.~V.}\ \bibnamefont {Sivak}}, \bibinfo {author}
  {\bibfnamefont {S.}~\bibnamefont {Puri}}, \bibinfo {author} {\bibfnamefont
  {A.}~\bibnamefont {Grimm}}, \bibinfo {author} {\bibfnamefont
  {L.}~\bibnamefont {Frunzio}}, \bibinfo {author} {\bibfnamefont
  {S.}~\bibnamefont {Shankar}},\ and\ \bibinfo {author} {\bibfnamefont {M.~H.}\
  \bibnamefont {Devoret}},\ }\bibfield  {title} {\bibinfo {title} {Gated
  {Conditional} {Displacement} {Readout} of {Superconducting} {Qubits}},\
  }\href {https://doi.org/10.1103/PhysRevLett.122.080502} {\bibfield  {journal}
  {\bibinfo  {journal} {Physical Review Letters}\ }\textbf {\bibinfo {volume}
  {122}},\ \bibinfo {pages} {80502} (\bibinfo {year} {2019})},\ \bibinfo {note}
  {arXiv: 1809.06964 Publisher: American Physical Society}\BibitemShut
  {NoStop}%
\bibitem [{\citenamefont {Ryan}\ \emph {et~al.}(2015)\citenamefont {Ryan},
  \citenamefont {Johnson}, \citenamefont {Gambetta}, \citenamefont {Chow},
  \citenamefont {Da~Silva}, \citenamefont {Dial},\ and\ \citenamefont
  {Ohki}}]{ryan_tomography_2015}%
  \BibitemOpen
  \bibfield  {author} {\bibinfo {author} {\bibfnamefont {C.~A.}\ \bibnamefont
  {Ryan}}, \bibinfo {author} {\bibfnamefont {B.~R.}\ \bibnamefont {Johnson}},
  \bibinfo {author} {\bibfnamefont {J.~M.}\ \bibnamefont {Gambetta}}, \bibinfo
  {author} {\bibfnamefont {J.~M.}\ \bibnamefont {Chow}}, \bibinfo {author}
  {\bibfnamefont {M.~P.}\ \bibnamefont {Da~Silva}}, \bibinfo {author}
  {\bibfnamefont {O.~E.}\ \bibnamefont {Dial}},\ and\ \bibinfo {author}
  {\bibfnamefont {T.~A.}\ \bibnamefont {Ohki}},\ }\bibfield  {title} {\bibinfo
  {title} {Tomography via correlation of noisy measurement records},\ }\href
  {https://doi.org/10.1103/PhysRevA.91.022118} {\bibfield  {journal} {\bibinfo
  {journal} {Physical Review A}\ }\textbf {\bibinfo {volume} {91}},\ \bibinfo
  {pages} {022118} (\bibinfo {year} {2015})}\BibitemShut {NoStop}%
\end{thebibliography}%
\end{document}